\documentclass[authoryear,3p,times,final]{elsarticle}

\usepackage{amssymb,multicol}
\usepackage{amsmath,comment}

\usepackage{graphicx,algorithm,epstopdf,lineno}

\usepackage{color,url,float,epsf}
\usepackage{microtype,bm}
\usepackage[colorlinks=true,breaklinks=true,pdftex]{hyperref}

\newcommand{\bez}{B\'{e}zier\ }
\newcommand{\be}[1]{\begin{equation}\label{#1}}
\newcommand{\ee}{\end{equation}}
\newcommand{\proof}{{\noindent\bf Proof.\ }}
\newcommand{\bb}{{\boldsymbol{\rho}_1}}
\newcommand{\pp}{{\boldsymbol{\rho}_2}}

\newtheorem{thm}{Theorem}
\newdefinition{definition}{Definition}
\newdefinition{rmk}{Remark}
\newdefinition{cor}{Corollary}
\newdefinition{claim}{Claim}
\newdefinition{proposition}{Proposition}
\newtheorem{lem}{Lemma}

\newdefinition{example}{Example}

\journal{arXiv.org}
\date{}
\begin{document}

\begin{frontmatter}

\title{Testing tensor product B\'{e}zier surfaces for coincidence: A comprehensive solution}

\author[inst1]{Krassimira Vlachkova}
\cortext[mycorrespondingauthor]{Corresponding author}
\ead{krassivl@fmi.uni-sofia.bg}

\affiliation[inst1]{organization={Faculty of Mathematics and Informatics, Sofia University ``St. Kliment Ohridski"},
	            addressline={Blvd. James Bourchier 5},
                city={Sofia},
	            postcode={1164},
	            country={Bulgaria}}

\begin{abstract}
It is known that B\'{e}zier curves and surfaces may have multiple representations by different control polygons. The polygons may have different number of control points and may even be disjoint. Up to our knowledge, Pekerman et al. (2005) were the first to address the problem of testing two parametric polynomial curves for coincidence. Their approach is based on reduction of the input curves into canonical irreducible form. They claimed that
their approach can be extended for testing tensor product surfaces but gave no further detail.

In this paper we develop a new technique and provide a comprehensive solution to the problem of testing tensor product \bez surfaces for coincidence. In (Vlachkova, 2017) an algorithm for testing \bez curves was proposed based on subdivision. There a partial solution to the problem of testing tensor product \bez surfaces was presented. Namely, the case where the irreducible surfaces are of same degree $(n,m)$, $n,m\in\mathbb{N}$, was resolved under certain additional condition. The other cases where one of the surfaces is of degree $(n,m)$ and the other is of degree either $(n,n+m)$, or $(n+m,m)$, or $(n+m,n+m)$ remained open.

We have implemented our algorithm for testing tensor product \bez surfaces for coincidence using Mathematica package. Experimental results and their analysis are presented.

\end{abstract}

\pagebreak
\end{frontmatter}


\section{Introduction}

 Assume we are given the control polygons of two tensor product B\'{e}zier (TPB) surfaces that are generated by different sources, e.\,g. different algorithms or software packages. The two polygons may have different number of control points and may even be disjoint but nevertheless it is possible that they represent surfaces with coincidence. Here we consider the problem of finding whether two control polygons represent different surfaces or partially/entirely coincident surfaces. In the latter case  we need to determine their coincident part. By {\it coincident} surfaces we mean that they occupy same locus of points in $\mathbb{R}^3$ but they may be parameterized differently, i.\,e. they are {\it geometrically equivalent} as defined in \citep{DH}.

This problem arises in various applications where the two surfaces need to be stitched together so that the obtained new surface is continuous. The problem is important also for the intersection algorithms based on subdivision which do not work well if the surfaces have coincident part.

The problem of testing polynomial curves for coincidence received a considerable attention by many authors. Up to our knowledge, \citet{PSEK} were the first to address the problem of testing two parametric polynomial curves for coincidence. They represented the curves into irreducible form\footnote{Different representations of a polynomial curve may occur if it has been degree elevated and/or reparameterized by a composition with a polynomial. A curve is {\it irreducible} if it is not a result of a polynomial composition and has not been degree elevated. Checking curves for irreducibility is well understood, see e.\,g. \citep{BZ2,KL,G}. In our experiments we use the built-in functions in Mathematica package.}, tested them for coincidence and determined their
shared domain in the case where coincidence occurs. In \citep{BP} a general approach and result for comparing rational \bez curves based on their control polygons was proposed. Then, in a series of papers this approach for testing \bez curves for coincidence has been developed and discussed, see \citep{WZLP,S-R1,CMD,S-R2,CYM,S-R4,CM}.

The use of control polygons when comparing \bez curves is preferable over their reparametrization as proposed in \citep{PSEK}
due to stability and numerical issues, see \citep{S-R3,F1}.
In addition, the transformations involved become ill-conditioned for high degrees.
In \citep{Vlachkova} an algorithm for testing \bez curves for coincidence based on control polygons and subdivision was presented, analysed and experimentally tested.

\citet{PSEK} suggested in the concluding remarks that their approach for testing polynomial curves for coincidence can be extended to TPB surfaces. No further detail was given. \citet{Vlachkova} proposed an algorithm for testing TPB surfaces as a generalization of the algorithm for curves presented in the same paper. The algorithm is based on comparing the control polygons of the two surfaces.
TPB surface $S(u,v)$,
$0\leq u\leq 1$, $0\leq v\leq 1$,
is {\it irreducible} if all comprising B\'{e}zier curves in $u$ and $v$ directions are irreducible.
Two irreducible TPB surfaces can have coincident part either (i) if they are of same degree $(n,m)$, $n,m\in\mathbb{N}$; or (ii) if one is of degree $(n,m)$ and the other is of degree either $(n,n+m)$, or $(n+m,m)$, or $(n+m,n+m)$.
The algorithm in \citep{Vlachkova} works only for TPB surfaces of same degree $(n,m)$ with overlapping boundary curves, see Fig.~\ref{nm1}{\bf a.} All other cases remained open.

Here we apply a different approach and present a complete solution to the problem of comparing TPB surfaces for coincidence. First, we show that all cases where two irreducible TPB surfaces have coincidence can be reduced to two main cases: (i) the surfaces are of same degree; (ii) the surfaces are of degree $(n,m)$ and $(n+m,n+m)$, respectively. Then, we reduce the problem to solving a nonlinear system of equations of degree $n+m$. The number of the unknowns is four and eight in cases (i) and (ii), respectively. Finally, we propose a method for solving these systems. Based on their solutions, we decide whether the input surfaces are different or partially/entirely coincident. In the latter case we determine the control points of the coincident part  using the blossoming principle. We also derive sufficient geometric criteria for checking whether the surfaces are different.

We have implemented our approach using Mathematica package. We reformulate the arising nonlinear systems of high degree so that Mathematica finds correctly their solutions.
The experimental results are presented, analysed and visualized.

The paper is organized as follows. In Section~\ref{sec:1} we formulate the problem and consecutively resolve cases (i) and (ii) in Subsection~\ref{subsec:1} and Subsection~\ref{subsec:2}, respectively. In Section~\ref{sec:2} we discuss the implementation and present our experimental results. Summary and conclusions are presented in Section~\ref{sec:3}.

\section{Coincidence of TPB surfaces}
\label{sec:1}

 Tensor product \bez surface $S(u,v)$ of degree $(n,m)$ for $n,m\in\mathbb{N}$, and control points ${\bf p}_{ij}\in\mathbb{R}^3$, $i=0,\dots, n$, $j=0,\dots, m$ is defined by
 \begin{equation}\label{ee1}
S(u,v)=\sum_{i=0}^n\sum_{j=0}^m{\bf p}_{ij}B_i^n(u)B_j^m(v),
\end{equation}
where $0\leq u\leq 1$, $0\leq v\leq 1$, and  $B_i^n(u):={n\choose i} u^i(1-u)^{n-i}$ are the Bernstein polynomials. Hereafter we assume that the binomial coefficients ${n\choose i}=0$ if $i<0$ or $i>n$.

We denote by $L(S)$ the locus of points $(u,v,S(u,v))$ in $\mathbb{R}^3$ for $0\leq u\leq 1$, $0\leq v\leq 1$.
\begin{definition}
TPB surfaces $S_1$ and $S_2$ have {\it coincidence} if there exists TPB surface $S$ such that $L(S_k)\subseteq L(S)$, $k=1,2$. We distinguish the following three cases.
\begin{enumerate}[(i)]
\item if $L(S_1)\equiv L(S_2)$ then ${S}_1$ and ${S}_2$ are {\it coincident},
\item if $L(S_1)\cap L(S_2)\not=\emptyset$ then ${S}_1$ and ${S}_2$ have {\it coincident part},
\item if $L(S_1)\cap L(S_2)=\emptyset$ then ${S}_1$ and ${S}_2$ are {\it disjoint}.
\end{enumerate}
${S}_1$ and ${S}_2$ are different if they do not have coincidence.
\end{definition}

Recall that a \bez curve is irreducible if it is not a result of a polynomial composition and has not been degree elevated.
\begin{definition}
TPB surface $S(u,v)$
is {\it irreducible} if the B\'{e}zier curves ${C}^u_j(u)$, $j=0,\dots, m$ with control points $\{{\bf p}_{ij}\}_{i=0}^n$,  and ${C}^v_i(v)$, $i=0,\dots, n$ with control points $\{{\bf p}_{ij}\}_{j=0}^m$ are irreducible.
\end{definition}
For TPB surface $S$ with control points ${\bf p}_{ij}$ we denote
by $\boldsymbol{\rho}$, ${\boldsymbol{\rho}}^{1,0}$, and ${\boldsymbol{\rho}}^{0,1}$ the finite differences at point ${\bf p}_{00}$ of order $(n,m)$, $(n-1,m)$, and $(n,m-1)$ respectively,  defined by (see \citep[pp. 66, 256]{F})
\begin{align}
&\boldsymbol{\rho} :=\Delta^{n,m}{\bf p}_{00}=\sum_{i=0}^n\sum_{j=0}^m(-1)^{n+m-i-j}{n\choose i}{m\choose j}{\bf p}_{ij},\nonumber\\
&{\boldsymbol{\rho}}^{1,0} :=\Delta^{n-1,m}{\bf p}_{00}=\frac{1}{n}\sum_{i=0}^n\sum_{j=0}^m(-1)^{n+m-i-j}{n\choose i}{m\choose j}(i-n){\bf p}_{ij},\label{e02}\\
&{\boldsymbol{\rho}}^{0,1} :=\Delta^{n,m-1}{\bf p}_{00}=\frac{1}{m}\sum_{i=0}^n\sum_{j=0}^m(-1)^{n+m-i-j}{n\choose i}{m\choose j}(j-m){\bf p}_{ij}.\nonumber
\end{align}

\begin{rmk}\label{rmk5}
If $S$ is irreducible then $\boldsymbol{\rho}$ is non-collinear to both ${\boldsymbol{\rho}}^{1,0}$ and ${\boldsymbol{\rho}}^{0,1}$.
\end{rmk}

It is known that two irreducible TPB surfaces $S_1$ and $S_2$ of degrees $(n,m)$ and $(n_1,m_1)$, respectively, may have coincidence only if $(n_1,m_1)$ equals to one of the following: $(n,m)$, $(n,n+m)$, $(n+m,m)$, $(n+m,n+m)$, see \citep[pp.\,253]{F}.
We consider first the cases (i) $(n_1,m_1)=(n,m)$, $n,m\in\mathbb{N}$ and (ii) $(n_1,m_1)=(n+m,n+m)$. Then we show that the other two cases reduce to (ii).

The next proposition is shown in \citep{BP} and \citep{S-R1} for curves. It can be easily extended to case (i) and to case (ii) (see Theorem~1 in \citep{YZ}) as follows.
\begin{proposition}\label{proposition1}
\begin{enumerate}[(i)]
\item The irreducible TPB surfaces $S_1(u,v)$ and $S_2(s,t)$ of same degree $(n,m)$ have coincidence if and only if there exists affine transformation
\be{k1}
\varphi : \left\{\begin{array}{l}
u(s)=(1-s)a+sb,\\
v(t)=(1-t)c+td,
\end{array}
\right.
\ee
such that $S_2(s,t)=S_1(u(s),v(t))$ for $0\leq s\leq 1$, $0\leq t\leq 1$, see Fig.~\ref{Paper_domain}(i).
\item The irreducible TPB surfaces $S_1(u,v)$ and $S_2(s,t)$ of degrees $(n,m)$ and $(n+m,n+m)$, respectively, have coincidence if and only if there exists bilinear transformation
    \be{k5}
\psi : \left\{
\begin{array}{l}
u(s,t)=(1-s)(1-t)a_1+(1-s)td_1
+stc_1+s(1-t)b_1,\\
v(s,t)=(1-s)(1-t)a_2+(1-s)td_2
+stc_2+s(1-t)b_2
\end{array}
\right.
\ee

such that $S_2(s,t)=S_1(u(s,t),v(s,t))$ for $0\leq s\leq 1$, $0\leq t\leq 1$, see Fig.~\ref{Paper_domain}(ii).
\end{enumerate}
\end{proposition}

\subsection{Irreducible TPB surfaces of same degree $(n,m)$}
\label{subsec:1}

Let
$S_1(u,v)$
and
$S_2(s,t)$
be irreducible TPB surfaces of same degree $(n,m)$ defined as in (\ref{ee1}) with control points ${\bf p}^1_{ij}$ and ${\bf p}^2_{ij}$, respectively. Let the corresponding vectors ${\boldsymbol{\rho}}_k$, ${\boldsymbol{\rho}}^{1,0}_k$, and ${\boldsymbol{\rho}}^{0,1}_k$, $k=1,2$, be defined by (\ref{e02}). We assume that $S_1$ and $S_2$ have different control polygons. In the next lemma we derive necessary geometric conditions for $S_1$ and $S_2$ to have coincidence.

\begin{lem}\label{lem1}
If the irreducible TPB surfaces $S_1$ and $S_2$ of degree $(n,m)$ have coincidence then the following statements hold.
\begin{enumerate}[(i)]
\item $\bb$ and $\pp$ are collinear;
\item $\boldsymbol{\rho}_1$, $\boldsymbol{\rho}_2$, ${\boldsymbol{\rho}}_1^{1,0}$, and ${\boldsymbol{\rho}}_2^{1,0}$ are coplanar;
\item $\boldsymbol{\rho}_1$, $\boldsymbol{\rho}_2$, ${\boldsymbol{\rho}}_1^{0,1}$, and ${\boldsymbol{\rho}}_2^{0,1}$ are coplanar.
\end{enumerate}
\end{lem}

\begin{figure}[ht]
 	\begin{minipage}[b]{1.85in}
\hspace*{.5cm}
\includegraphics[width=1.5\textwidth]{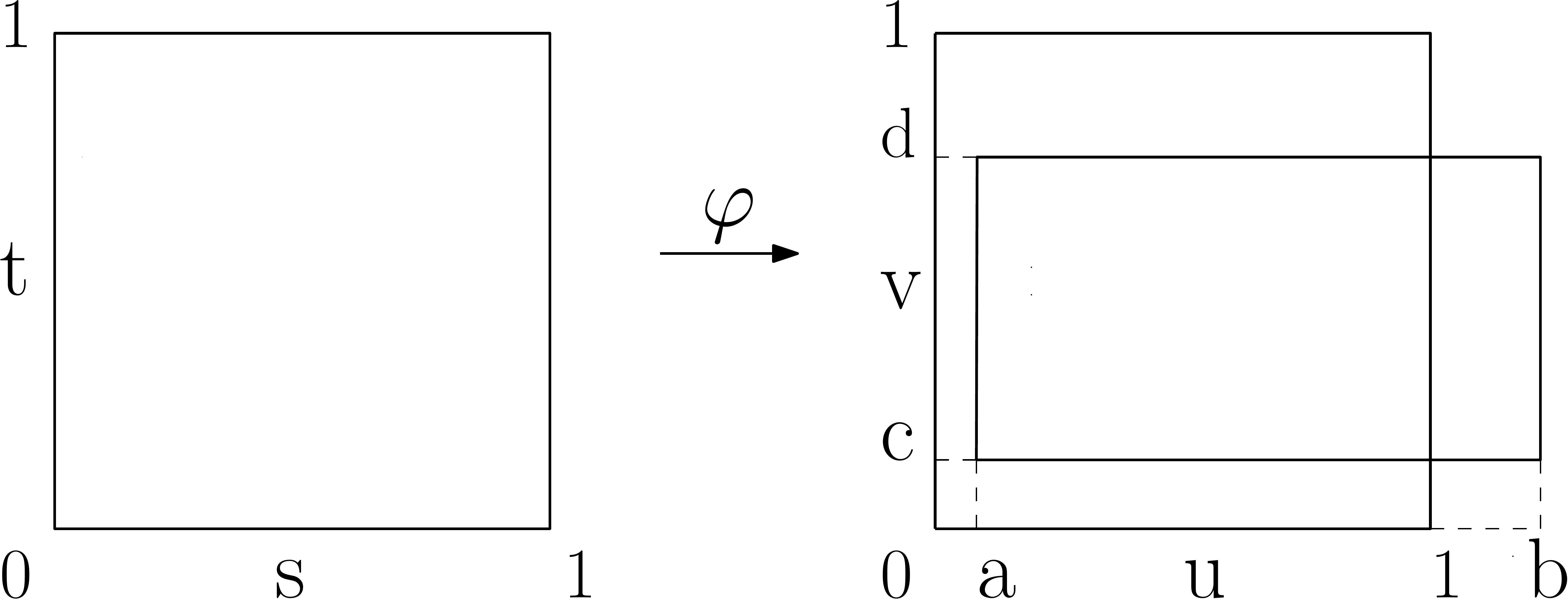}
\hspace*{4.cm}{\small {(i)}}
    \end{minipage}
       \hspace{2.5cm}
    ~~~~~~~~~
    \begin{minipage}[b]{1.85in}
    	\hspace*{.5cm}
\includegraphics[width=1.5\textwidth]{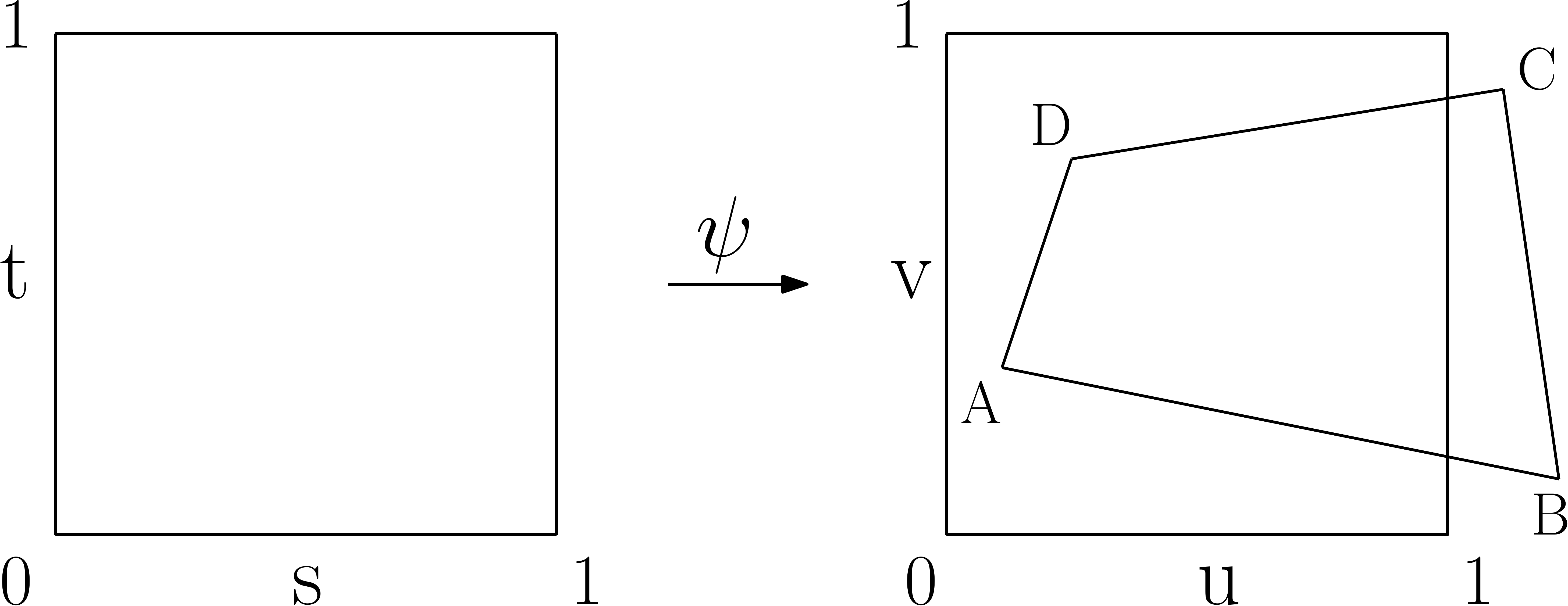}
\hspace*{4.cm}{\small {(ii)}}
        \end{minipage}
           \caption{\small {(i)} The domain $\{(u,v):a\leq u\leq b,\ c\leq v\leq d\}$ is an image of the domain
       $\{(s,t):0\leq s\leq 1,\ 0\leq t\leq 1\}$ under the affine transformation $u=u(s)=(1-s)a+sb$, $v=v(t)=(1-t)c+td$.  {(ii)} The convex quadrilateral $ABCD$ is an image of the domain $\{(s,t)\,:\, 0\leq s\leq 1,\ 0\leq t\leq 1\}$ under a bilinear transformation.}\label{Paper_domain}
        \end{figure}

\proof Assume that $S_1$ and $S_2$ have coincidence.
 According to statement (i) of Proposition\,\ref{proposition1} there are four numbers $a,b,c,d\in\mathbb{R}$ such that the domain $\{(u,v):a\leq u\leq b,\ c\leq v\leq d\}$ is an image of the domain $\{(s,t):0\leq s\leq 1,\ 0\leq t\leq 1\}$ under the affine transformation (\ref{k1}) (see Fig.~\ref{Paper_domain}(i)) and $S_2(s,t)=S_1(u(s),v(t))$.

We have

\begin{equation}\label{e1}
\sum_{i=0}^{n}\sum_{j=0}^{m}{\bf p}^2_{ij}B_i^n(s)B_j^m(t)=\sum_{i=0}^{n}\sum_{j=0}^{m}{\bf p}^1_{ij}B_i^n((1-s)a+sb)B_j^m((1-t)c+td),\  0\leq s\leq 1,\ 0\leq t\leq 1.
\end{equation}

We take $\frac{\partial^{n+m}}{\partial s^n\partial t^m}$ derivative in (\ref{e1}) and obtain
$$n!m!\pp=n!m!(b-a)^n(d-c)^m\bb.$$ Hence, vectors $\bb$ and $\pp$ are collinear with $\pp=\kappa\bb$ where we denoted
\be{e10}
\kappa=(b-a)^n(d-c)^m.
\ee
This proves statement (i).

Since $\bb$ and $\pp$ are the coefficients of $u^nv^m$ and $s^nt^m$ in $S_1(u,v)$ and $S_2(s,t)$, respectively, and $S_1$ and $S_2$ are irreducible then $\bb\not={\bf 0}$ and $\pp\not={\bf 0}$. So the number $\kappa$ is determined as the ratio of any two corresponding nonzero coordinates of $\pp$ and $\bb$.

Next, we take $\frac{\partial^{n+m-1}}{\partial s^{n-1}\partial t^m}$ derivative in (\ref{e1}) and obtain
\begin{equation*}
\begin{split}
m!(n-1)!\sum_{i=0}^n\sum_{j=0}^m(-1)^{n+m-i-j}{n\choose i}{m\choose j}(ns-n+i){\bf p}^2_{ij}=\\
m!(n-1)!(b-a)^{n-1}(d-c)^m\sum_{i=0}^n\sum_{j=0}^m(-1)^{n+m-i-j}{n\choose i}{m\choose j}(n(s(b-a)+a)-n+i){\bf p}^1_{ij},\\
0\leq s\leq 1,
\end{split}
\end{equation*}
which for $s=0$ implies
\be{e4}
{\boldsymbol{\rho}_2^{1,0}}=(b-a)^{n-1}(d-c)^m(a\bb+{\boldsymbol{\rho}_1^{1,0}}).
\ee

Hence, vectors $\boldsymbol{\rho}_1$, ${\boldsymbol{\rho}}_1^{1,0}$, and ${\boldsymbol{\rho}}_2^{1,0}$ are coplanar and (ii) follows from (i).

Similarly, we take $\frac{\partial^{n+m-1}}{\partial s^{n}\partial t^{m-1}}$ derivative in (\ref{e1})
and for $s=0$  obtain
\be{e6}
{\boldsymbol{\rho}_2^{0,1}}=(b-a)^{n}(d-c)^{m-1}(c\bb+{\boldsymbol{\rho}_1^{0,1}}).
\ee

Therefore vectors $\boldsymbol{\rho}_1$, ${\boldsymbol{\rho}}_1^{0,1}$, and ${\boldsymbol{\rho}}_2^{0,1}$ are coplanar and (iii) follows from (i).
\hfill $\Box$

\begin{rmk}\label{rmk4}
Surface $S_2$ can be considered as obtained from surface $S_1$ by subdivision with respect to $u$ at $a$ and $b$, and with respect to $v$ at $c$ and $d$.
\end{rmk}

Next we obtain necessary and sufficient conditions for surfaces $S_1$ and $S_2$ to have coincidence. In the proof of Lemma~\ref{lem1} we have shown that if $S_1$ and $S_2$ have coincidence then there exist numbers $a,b,c,d$ defining affine transformation (\ref{k1}) and satisfying the system
\be{k2}
\left|
\begin{array}{l}
{\boldsymbol{\rho}_2^{1,0}}=(b-a)^{n-1}(d-c)^m(a\bb+{\boldsymbol{\rho}_1^{1,0}}),\\
{\boldsymbol{\rho}_2^{0,1}}=(b-a)^{n}(d-c)^{m-1}(c\bb+{\boldsymbol{\rho}_1^{0,1}}),\\
\kappa=(b-a)^n(d-c)^m,
\end{array}
\right.
\ee
where $\kappa$ is determined from $\boldsymbol{\rho}_2=\kappa\boldsymbol{\rho}_1$.
\begin{lem}\label{lem3}
If system (\ref{k2}) is consistent then it has a unique solution.
\end{lem}

\proof We multiply the first vector equation in (\ref{k2}) by $b-a$ and the second one by $d-c$, use $\boldsymbol{\rho}_2=\kappa\boldsymbol{\rho}_1$, and obtain
\be{e11}
b{\boldsymbol{\rho}_2^{1,0}}-a(\pp+{\boldsymbol{\rho}_2^{1,0}})=
\kappa{\boldsymbol{\rho}_1^{1,0}},
\ee
\be{e12}
d{\boldsymbol{\rho}_2^{0,1}}-c(\pp+{\boldsymbol{\rho}_2^{0,1}})=
\kappa{\boldsymbol{\rho}_1^{0,1}}.
\ee
Hence, (\ref{k2}) is equivalent to the following system
\be{k10}
\left|
\begin{array}{l}
b{\boldsymbol{\rho}_2^{1,0}}-a(\pp+{\boldsymbol{\rho}_2^{1,0}})=
\kappa{\boldsymbol{\rho}_1^{1,0}},\\
d{\boldsymbol{\rho}_2^{0,1}}-c(\pp+{\boldsymbol{\rho}_2^{0,1}})=
\kappa{\boldsymbol{\rho}_1^{0,1}},\\
\kappa=(b-a)^n(d-c)^m,
\end{array}
\right.
\ee
Each of the two vector equations in (\ref{k10}) is equivalent to a linear system of three equations of the unknowns $a$,\,$b$, and $c$,\,$d$, respectively, for each of the three vector coordinates. By Remark\,\ref{rmk5} the ranks of the matrices of these systems are greater than one. Hence the systems have either unique, or no solution. Therefore, if system (\ref{k10}) is consistent then it has a unique solution.
\hfill $\Box$

The following theorem holds.
\begin{thm}\label{thm1}
The irreducible TPB surfaces $S_1$ and $S_2$ of degree $(n,m)$ have coincidence if and only if system (\ref{k10}) is consistent and the control polygon of the surface corresponding to the unique solution to (\ref{k10}) coincide with the control polygon of $S_2$ (up to eight different enumerations of the control polygons).
\end{thm}

\proof $\Rightarrow$ Let $S_1$ and $S_2$ have coincidence. Then according to statement (i) of Proposition\,\ref{proposition1} there exist four numbers $a,b,c,d\in\mathbb{R}$ defining transformation $\varphi$ which are a solution to system (\ref{k2}) and $S_2(s,t)=S_1(\varphi)$. According to Lemma~\ref{lem3}, the equivalent to (\ref{k2}) system (\ref{k10}) has a unique solution $(a,b,c,d)$. The control polygon of the surface corresponding to this solution coincides with the control polygon of $S_2$.

$\Leftarrow$ Let system (\ref{k10}) be consistent. Then, according to Lemma~\ref{lem3}, it has a unique solution. Since the control polygon of the surface $S$ corresponding to this solution coincide with the control polygon of $S_2$ (up to eight different enumerations of the control polygons) then $S$ and $S_2$ coincide according to Theorem~1 in \citep{Vlachkova}.
\hfill $\Box$
\vspace{2ex}

We continue by providing an efficient approach for testing $S_1$ and $S_2$ for coincidence. First, we consider the two linear systems (\ref{e11}) and (\ref{e12}) which have either unique or no solution. Clearly if any of them has no solution then by Proposition~\ref{proposition1} no transformation $\varphi$ exists and $S_1$ and $S_2$ are different. If both systems have unique solutions, say $(a^*,b^*)$, $(c^*,d^*)$, then we need to check if they satisfy $(b^*-a^*)^n(d^*-c^*)^m=\kappa$. If they do not, then system (\ref{k10}) is inconsistent and $S_1$ and $S_2$ are different. Otherwise, following Theorem~\ref{thm1}, we need to compute the control polygon of surface $S^*$ corresponding to $(a^*,b^*,c^*,d^*)$ and to check if it coincides with the control polygon of $S_2$. If these polygons coincide (up to eight different enumerations of the control points) then $S_1$ and $S_2$ have coincidence, otherwise they are different. We compute the control points of $S^*$ using the blossoming principle, see \citep{RG}. In \citep[p. 339]{RG} and \citep{YZ} it is pointed out that for any polynomial surface patch $S(u,v)=\sum_{i=0}^n\sum_{j=0}^m{\bf c}_{ij}u^iv^j$ of degree $(n,m)$ defined for $a\leq u\leq b$ and $c\leq v\leq d$, the B\'{e}zier control points ${\bf p}_{\nu\mu}$, $\nu=0,\dots ,n$, $\mu=0,\dots ,m$, of this surface patch are
\be{e14}
{\bf p}_{\nu\mu}=b^{\Box} (\underbrace{a,\dots ,a}_\text{$\nu$},\underbrace{b,\dots ,b}_\text{$n-\nu$},\underbrace{c,\dots ,c}_\text{$\mu$},\underbrace{d,\dots ,d}_\text{$m-\mu$}),
\ee
where
$b^{\Box}(u_1,\dots ,u_n,v_1,\dots ,v_m)=\sum_{i=0}^n\sum_{j=0}^m{\bf c}_{ij}b_{ij}^{\Box} (u_1,\dots ,u_n,v_1,\dots ,v_m)$
is the blossom of $S(u,v)$, and
\begin{equation}\label{e225}
b_{ij}^{\Box} (u_1,\dots ,u_n,v_1,\dots ,v_m)
=\sum_{\{\alpha_1,\dots ,\alpha_i\}\subseteq \{1,\dots ,n\}}\frac{u_{\alpha_1}\dots u_{\alpha_i}}{\binom{n}{i}} \sum_{\{\beta_1,\dots ,\beta_j\}\subseteq \{1,\dots ,m\}}\frac{v_{\beta_1}\dots v_{\beta_j}}{\binom{m}{j}}
\end{equation}
 is the blossom  of the monomial $u^iv^j$. In the case where either $i=0$, or $j=0$, the corresponding sum in (\ref{e225}) equals 1, e.\,g. $b_{00}^{\Box}=1$. In the next corollary we present (\ref{e14}) and (\ref{e225}) in an equivalent closed form that is more suitable for computations.
 \begin{cor}\label{cor1}
 B\'{e}zier control points ${\bf p}_{\nu\mu}$, $\nu=0,\dots ,n$, $\mu=0,\dots ,m$, defined by (\ref{e14}) are
 \be{e226}
 {\bf p}_{\nu\mu}=\sum_{i=0}^n\sum_{j=0}^m\frac{{\bf c}_{ij}}{\binom{n}{i}\binom{m}{j}}\sum_{k=\max (0,i+\nu-n)}^{\min (i,\nu)}\ \sum_{r=\max (0,j+\mu-m)}^{\min (j,\mu)} \binom{\nu}{k}\binom{n-\nu}{i-k}\binom{\mu}{r}\binom{m-\mu}{j-r}a^kb^{i-k}c^rd^{j-r}.
 \ee
\end{cor}
\vspace{2ex}

We outline our procedure for testing $S_1$ and $S_2$ for coincidence in algorithmic form below.
\begin{algorithm}[ht]
\caption{Testing two irreducible TPB surfaces of degree $(n,m)$ for coincidence}\label{alg1}
{\small
\begin{tabular}{rl}
  {\sl Input:}  & Irreducible TPB surfaces $S_1$ and $S_2$ of degree $(n,m)$ given by their control polygons\\
    {\sl Output:} & \ \ (i) $S_1$ and $S_2$ are different;\\
                & \ (ii) $S_1$ and $S_2$ are disjoint;\\
                & (iii) $S_1$ and $S_2$ have coincident part. Report its control points.\\
  {\sl Step~1.} &Compute vectors $\boldsymbol{\rho}_i$, ${\boldsymbol{\rho}}_i^{1,0}$, and ${\boldsymbol{\rho}}_i^{0,1}$, $i=1,2$.\\
  {\sl Step~2.} & Check the conditions of Lemma~\ref{lem1}. \\
   & {\sl 2.1.} {\bf If} $\boldsymbol{\rho}_1$, $\boldsymbol{\rho}_2$ are non-collinear\\
  &\hspace{2mm}\qquad {\bf then} return (i);\\
  &\hspace{2mm}\quad \qquad {\bf else} compute $\kappa$ such that $\boldsymbol{\rho}_2=\kappa\boldsymbol{\rho}_1$.\\
  &{\sl 2.2.} {\bf If} either $\boldsymbol{\rho}_1$, $\boldsymbol{\rho}_2$, ${\boldsymbol{\rho}}_1^{1,0}$, ${\boldsymbol{\rho}}_2^{1,0}$, or $\boldsymbol{\rho}_1$, $\boldsymbol{\rho}_2$, ${\boldsymbol{\rho}}_1^{0,1}$, ${\boldsymbol{\rho}}_2^{0,1}$ are non-coplanar\\
  &\hspace{2mm}\qquad {\bf then} return (i);\\
   &\hspace{2mm}\quad\qquad {\bf else} system (\ref{k10}) has either unique, or no solution.\\
    {\sl Step~3.} & Solve linear systems (\ref{e11}) and (\ref{e12}).\\
     & {\bf If} any of them is inconsistent\\
     &\quad  {\bf then} return (i);\\
     &\qquad {\bf else} denote their unique solutions by $(a^*,b^*)$ and $(c^*,d^*)$. \\
  {\sl Step~4.} & {\bf If} $(b^*-a^*)^n(d^*-c^*)^m\not=\kappa$\\
   &\quad {\bf then} return (i);\\
   &\qquad  {\bf else} system (\ref{k2}) is consistent with unique solution $(a^*,b^*,c^*,d^*)$.\\
  {\sl Step~5.} & Compute the control polygon of the transformed TPB surface\\
   &$S^*(s,t)=S_1(u(\varphi(s,t),v(\varphi(s,t))$ using (\ref{e226}) and compare it to the control polygon of $S_2$.\\
    & {\bf If} they coincide (up to eight different enumerations)\\
    &\quad  {\bf then} $S_1$ and $S_2$ have coincidence;\\
    &\qquad {\bf else} return (i).\\
  {\sl Step~6.} & Compute the shared domain $G$ of $S^*$ and $S$.\\
   & {\bf If} $D=\emptyset$\\
   &\quad {\bf then} return (ii);\\
   &\qquad {\bf else} compute the control points of the coincident part using (\ref{e226}) and return (iii).
     \end{tabular}
}

\end{algorithm}

\subsection{Irreducible B\'{e}zier surfaces of degrees $(n,m)$ and $(n+m,n+m)$}
\label{subsec:2}

Let $S_1(u,v)=\sum_{i=0}^n\sum_{j=0}^m{\bf p}^1_{ij}B_i^n(u)B_j^m(v)$ defined for $0\leq u\leq 1$, $0\leq v\leq 1$, and
$S_2(s,t)=\sum_{i=0}^{n+m}\sum_{j=0}^{n+m}{\bf p}^2_{ij}B_i^{n+m}(s)B_j^{n+m}(t)$ defined for $0\leq s\leq 1$, $0\leq t\leq 1$, be irreducible TPB surfaces.
For the four boundary curves of $S_2$ we denote by ${\boldsymbol{\delta}_k}$ and ${\boldsymbol{\delta}_k^1}$, $k=1,\dots ,4$ the following finite differences
\begin{align*}
&{\boldsymbol{\delta}_1}:=\Delta^{n+m,0}{\bf p}^2_{00}=\sum_{i=0}^{n+m}(-1)^{n+m-i}{{n+m}\choose i}{\bf p}^2_{i0},\\
&{\boldsymbol{\delta}_2}:=\Delta^{0,n+m}{\bf p}^2_{00}=\sum_{i=0}^{n+m}(-1)^{n+m-i}{{n+m}\choose i}{\bf p}^2_{0i},\\
&{\boldsymbol{\delta}_3}:=\Delta^{n+m,0}{\bf p}^2_{0,n+m}=\sum_{i=0}^{n+m}(-1)^{n+m-i}{{n+m}\choose i}{\bf p}^2_{i,n+m},\\
&{\boldsymbol{\delta}_4}:=\Delta^{0,n+m}{\bf p}^2_{n+m,0}=\sum_{i=0}^{n+m}(-1)^{n+m-i}{{n+m}\choose i}{\bf p}^2_{n+m,i},\\
&{\boldsymbol{\delta}_1^1}:=(n+m)\Delta^{n+m-1,0}{\bf p}_{00}^2=\sum_{i=0}^{n+m}(-1)^{n+m-i}{{n+m}\choose i}(i-n-m){\bf p}^2_{i0},\\
&{\boldsymbol{\delta}_2^1}:=(n+m)\Delta^{0,n+m-1}{\bf p}_{00}^2=\sum_{i=0}^{n+m}(-1)^{n+m-i}{{n+m}\choose i}(i-n-m){\bf p}^2_{0i},\\
&{\boldsymbol{\delta}_3^1}:=(n+m)\Delta^{n+m-1,0}{\bf p}_{0,n+m}^2=\sum_{i=0}^{n+m}(-1)^{n+m-i}{{n+m}\choose i}(i-n-m){\bf p}^2_{i,n+m},\\
&{\boldsymbol{\delta}_4^1}:=(n+m)\Delta^{0,n+m-1}{\bf p}_{n+m,0}^2=\sum_{i=0}^{n+m}(-1)^{n+m-i}{{n+m}\choose i}(i-n-m){\bf p}^2_{n+m,i}.
\end{align*}

The finite differences $\boldsymbol{\rho}$, $\boldsymbol{\rho}^{1,0}$, and $\boldsymbol{\rho}^{0,1}$ for $S_1$ are defined by (\ref{e02}).

In the next lemma we derive necessary geometric conditions for $S_1$ and $S_2$ to have coincidence.

\begin{lem}\label{lem2}
If the irreducible TPB surfaces $S_1$ and $S_2$ of degrees $(n,m)$ and $(n+m,n+m)$, respectively, have coincidence then the five vectors $\boldsymbol{\rho}$, and $\boldsymbol{\delta}_i$, $i=1,\dots ,4$, are collinear.
        \end{lem}

        \proof Assume that $S_1$ and $S_2$ have coincidence. Since $S_1$ is of degree $(n,m)$ and $S_2$ is of degree $(n+m,n+m)$ then, according to statement (ii) of Proposition\,\ref{proposition1},  there exist convex quadrilateral $ABCD$ with vertices $A(a_1,a_2)$, $B(b_1,b_2)$, $C(c_1,c_2)$, $D(d_1,d_2)$ which is an image of the domain $\{(s,t)\,:\, 0\leq s\leq 1,\ 0\leq t\leq 1\}$ under the bilinear transformation (\ref{k5}) (see Fig.~\ref{Paper_domain}(ii)) and $S_2(s,t)=S_1(u(s,t),v(s,t))$. Hence, for $0\leq s\leq 1$, $0\leq t\leq 1$ we have
\be{e16}
\sum_{i=0}^{n}\sum_{j=0}^{m}{\bf p}^1_{ij}B_i^n(u)B_j^m(v)=\sum_{i=0}^{n+m}\sum_{j=0}^{n+m}{\bf p}^2_{ij}B_i^{n+m}(s)B_j^{n+m}(t),
\ee
where $u$ and $v$ are defined by (\ref{k5}).

The image of the boundary segment $\{(s,t):0\leq s\leq 1,\ t=0\}$  under the bilinear transformation (\ref{k5}) is the segment $AB$ and we have $u=(1-s)a_1+sb_1$, $v=(1-s)a_2+sb_2$. Hence, from (\ref{e16}) it follows
\begin{equation}\label{e17}
\sum_{i=0}^{n}\sum_{j=0}^{m}{\bf p}^1_{ij}B_i^n((1-s)a_1+sb_1)B_j^m((1-s)a_2+sb_2)
=\sum_{i=0}^{n+m}{\bf p}^2_{i0}B_i^{n+m}(s),\ 0\leq s\leq 1.
\end{equation}
After differentiation of (\ref{e17}) $n+m$ times we obtain
\be{e19}
(b_1-a_1)^n(b_2-a_2)^m\boldsymbol{\rho} =\boldsymbol{\delta}_1.
\ee
Hence, vectors $\boldsymbol{\rho}$ and $\boldsymbol{\delta}_1$ are collinear with $\kappa_1\boldsymbol{\rho} =\boldsymbol{\delta}_1$ where we denoted
\be{e019}
\kappa_1=(b_1-a_1)^n(b_2-a_2)^m.
\ee

Since $\boldsymbol{\rho}$ and $\boldsymbol{\delta}_1$ are the coefficients of $u^nv^m$ in $S_1(u,v)$ and $s^{n+m}$ in $S_2(s,0)$, respectively, and $S_1$ and $S_2$ are irreducible then $\boldsymbol{\rho}\not={\bf 0}$ and $\boldsymbol{\delta}_1\not={\bf 0}$. So the number $\kappa_1$ is determined as the ratio of any two corresponding nonzero coordinates of $\boldsymbol{\delta}_1$ and $\boldsymbol{\rho}$.

For the remaining three boundary segments we obtain, analogously to (\ref{e19}),
\begin{eqnarray}
&&(d_1-a_1)^n(d_2-a_2)^m\boldsymbol{\rho}=\boldsymbol{\delta}_2 ,\nonumber\\
&&(c_1-d_1)^n(c_2-d_2)^m\boldsymbol{\rho}=\boldsymbol{\delta}_3 ,\label{e20}\\
&&(c_1-b_1)^n(c_2-b_2)^m\boldsymbol{\rho}=\boldsymbol{\delta}_4 .\nonumber
\end{eqnarray}

Hence, vectors $\boldsymbol{\rho}$, $\boldsymbol{\delta}_2$, $\boldsymbol{\delta}_3$, and $\boldsymbol{\delta}_4$ are collinear with $\kappa_i\boldsymbol{\rho} =\boldsymbol{\delta}_i$, $i=2,3,4$, where we denoted
\begin{eqnarray}
&&\kappa_2=(d_1-a_1)^n(d_2-a_2)^m,\nonumber\\
&&\kappa_3=(c_1-d_1)^n(c_2-d_2)^m,\label{k16}\\
&&\kappa_4=(c_1-b_1)^n(c_2-b_2)^m.\nonumber
\end{eqnarray}
Similarly to $\kappa_1$, the numbers $\kappa_i$ for $i=2,3,4$ are determined as the ratio of any two corresponding nonzero coordinates of $\boldsymbol{\delta}_i$ and $\boldsymbol{\rho}$, respectively.
\hfill $\Box$
\vspace{2ex}

Next we obtain a necessary and sufficient conditions for surfaces $S_1$ and $S_2$ to have coincidence.
We differentiate (\ref{e17}) $n+m-1$ times and obtain for $s=0$
$$
(b_1-a_1)^{n-1}(b_2-a_2)^{m-1}\big(n\boldsymbol{\rho}^{1,0}(b_2-a_2)+m\boldsymbol{\rho}^{0,1}(b_1-a_1)
+n\boldsymbol{\rho}a_1(b_2-a_2)
+m\boldsymbol{\rho}a_2(b_1-a_1)\big)=\boldsymbol{\delta}_1^1.
$$
Similarly, for the remaining three boundary segments we obtain
$$
\begin{array}{r}
(d_1-a_1)^{n-1}(d_2-a_2)^{m-1}\big(n\boldsymbol{\rho}^{1,0}(d_2-a_2)+m\boldsymbol{\rho}^{0,1}(d_1-a_1)+n\boldsymbol{\rho}a_1(d_2-a_2)
+m\boldsymbol{\rho}a_2(d_1-a_1)\big)=\boldsymbol{\delta}_2^1,\\[.5ex]
(c_1-d_1)^{n-1}(c_2-d_2)^{m-1}\big(n\boldsymbol{\rho}^{1,0}(c_2-d_2)+m\boldsymbol{\rho}^{0,1}(c_1-d_1)+n\boldsymbol{\rho}d_1(c_2-d_2)
+m\boldsymbol{\rho}d_2(c_1-d_1)\big)=\boldsymbol{\delta}_3^1,\\[.5ex]
(c_1-b_1)^{n-1}(c_2-b_2)^{m-1}\big(n\boldsymbol{\rho}^{1,0}(c_2-b_2)+m\boldsymbol{\rho}^{0,1}(c_1-b_1)+n\boldsymbol{\rho}b_1(c_2-b_2)
+m\boldsymbol{\rho}b_2(c_1-b_1)\big)=\boldsymbol{\delta}_4^1.
\end{array}
$$
Therefore, if $S_1$ and $S_2$ have coincidence then the eight numbers $a_i,b_i,c_i,d_i$, $i=1,2$, defining bilinear transformation (\ref{k5}) satisfy the system
\be{k6}
\left|
\begin{array}{l}
(b_1-a_1)^{n-1}(b_2-a_2)^{m-1}(n\boldsymbol{\rho}^{1,0}(b_2-a_2)+m\boldsymbol{\rho}^{0,1}(b_1-a_1)+n\boldsymbol{\rho}a_1(b_2-a_2)
+m\boldsymbol{\rho}a_2(b_1-a_1))=\boldsymbol{\delta}_1^1,\\[.5ex]
(d_1-a_1)^{n-1}(d_2-a_2)^{m-1}(n\boldsymbol{\rho}^{1,0}(d_2-a_2)+m\boldsymbol{\rho}^{0,1}(d_1-a_1)+n\boldsymbol{\rho}a_1(d_2-a_2)
+m\boldsymbol{\rho}a_2(d_1-a_1))=\boldsymbol{\delta}_2^1,\\[.5ex]
(c_1-d_1)^{n-1}(c_2-d_2)^{m-1}(n\boldsymbol{\rho}^{1,0}(c_2-d_2)+m\boldsymbol{\rho}^{0,1}(c_1-d_1)+n\boldsymbol{\rho}d_1(c_2-d_2)
+m\boldsymbol{\rho}d_2(c_1-d_1))=\boldsymbol{\delta}_3^1,\\[.5ex]
(c_1-b_1)^{n-1}(c_2-b_2)^{m-1}(n\boldsymbol{\rho}^{1,0}(c_2-b_2)+m\boldsymbol{\rho}^{0,1}(c_1-b_1)+n\boldsymbol{\rho}b_1(c_2-b_2)
+m\boldsymbol{\rho}b_2(c_1-b_1))=\boldsymbol{\delta}_4^1,\\[.5ex]
\kappa_1=(b_1-a_1)^n(b_2-a_2)^m,\\[.5ex]
\kappa_2=(d_1-a_1)^n(d_2-a_2)^m,\\[.5ex]
\kappa_3=(c_1-d_1)^n(c_2-d_2)^m,\\[.5ex]
\kappa_4=(c_1-b_1)^n(c_2-b_2)^m,
\end{array}
\right.
\ee
where $\kappa_i$ is determined from $\kappa_i\boldsymbol{\rho} =\boldsymbol{\delta}_i$, $i=1,\dots ,4$.

Straightforward application of Mathematica packages and build-in functions doesn't yield solutions to system (\ref{k6}) efficiently. Hence, it is important to develop a method to simplify and solve it.
\begin{lem}\label{lem4}
If system (\ref{k6}) is consistent and rank($\boldsymbol{\rho},\boldsymbol{\rho}^{1,0},\boldsymbol{\rho}^{0,1}$)=3 then it has a unique solution. If rank($\boldsymbol{\rho},\boldsymbol{\rho}^{1,0},\boldsymbol{\rho}^{0,1}$)=2 then it 
has at most two solutions.
\end{lem}

\proof
We consider the first equation of system (\ref{k6}), denote $x:=b_1-a_1$, $y:=b_2-a_2$, and obtain
\be{e22}
x^{n-1}y^{m-1}(n{\boldsymbol{\rho}^{1,0}}y+m{\boldsymbol{\rho}^{0,1}}x+n\boldsymbol{\rho} a_1y+m\boldsymbol{\rho} a_2x)=\boldsymbol{\delta}_1^1.
\ee

Vector equation (\ref{e22}) is equivalent to the following system
\be{e23}
\left|
\begin{array}{l}
\kappa_1(n\rho^{1,0}_1y+m\rho^{0,1}_1x+n{{\rho}_1} a_1y+m{{\rho}_1} a_2x)=\delta_1xy,\\[1ex]
\kappa_1(n\rho^{1,0}_2y+m\rho^{0,1}_2x+n{{\rho}_2} a_1y+m{{\rho}_2} a_2x)=\delta_2xy,\\[1ex]
\kappa_1(n\rho^{1,0}_3y+m\rho^{0,1}_3x+n{{\rho}_3} a_1y+m{{\rho}_3} a_2x)=\delta_3xy,
\end{array}
\right.
\ee
where ${\boldsymbol{\rho}^{1,0}}=(\rho^{1,0}_1,\rho^{1,0}_2,\rho^{1,0}_3)$, ${\boldsymbol{\rho}^{0,1}}=(\rho^{0,1}_1,\rho^{0,1}_2,\rho^{0,1}_3)$, $\boldsymbol{\rho}=({\rho}_1,{\rho}_2,{\rho}_3)$, and 
$\boldsymbol{\delta}_1^1=(\delta_1,\delta_2,\delta_3)$.

To solve (\ref{e23}) we consider two cases according to the rank of the matrix $M$, where
\be{m1}
M=\left(
\begin{array}{cccc}
n\rho^{1,0}_1&m\rho^{0,1}_1&n{\rho}_{1}&m{\rho}_1\\
n\rho^{1,0}_2&m\rho^{0,1}_2&n{\rho}_{2}&m{\rho}_2\\
n\rho^{1,0}_3&m\rho^{0,1}_3&n{\rho}_{3}&m{\rho}_3
\end{array}
\right).
\ee

By Remark 1 the rank of M is greater that 1. Further on, $\alpha$, $\beta$, $C$, $\alpha_i$, $\beta_i$, $C_i$, $i=1,2$ denote real constants that depend on the input data only, more precisely on $\boldsymbol{\rho}^{1,0}$, $\boldsymbol{\rho}^{0,1}$, $\boldsymbol{\rho}$, and $\boldsymbol{\delta}_1^1$.

{\sf Case 1}. rank(M)=3

Since $\boldsymbol{\rho}$ is nonzero vector then some of its coordinates, say ${\rho}_1$, is nonzero.
We eliminate $a_1y$ and $a_2x$ from the last two equations of (\ref{e23}) by multiplying the first equation by $-{\rho}_2/{{\rho}_1}$ and $-{\rho}_3/{{\rho}_1}$ consecutively and adding it to the second and third equations, respectively.
We obtain a system of the following type
\be{e25}
\left|
\begin{array}{l}
\alpha_1x+\beta_1y=C_1xy,\\
\alpha_2x+\beta_2y=C_2xy,
\end{array}
\right.
\ee
which has a unique solution $(x,y)$.

We solve in an analogous way the remaining three vector equations of (\ref{k6}) with respect to the unknowns $d_1-a_1$ and $d_2-a_2$; $c_1-d_1$ and $c_2-d_2$; $c_1-b_1$ and $c_2-b_2$, respectively. Note that the corresponding three equivalent systems
have same matrix $M$ as system (\ref{e23}) and differ by their right sides only. Hence, each of them has also a unique solution which can be found straightforwardly. Therefore, if system (\ref{k6}) is consistent then it has a unique solution.

{\sf Case 2.} rank(M)=2

In this case, (\ref{e23}) has two linearly independent equations.
 Similar to Case 1, since the coefficients of $a_1y$ and $a_2x$ are in ratio $n:m$, then by multiplying one of these equations by a suitable constant and adding it to the other equation we exclude $a_1y$ and $a_2x$ and obtain one equation of the following type

 \be{e26}
\alpha x+\beta y=Cxy.
\ee
So we have to solve the following system
\be{e325}
\left|
\begin{array}{l}
	x^ny^m=\kappa_1,\\
	\alpha x+\beta y=Cxy.
\end{array}
\right.
\ee
\begin{claim}\label{c1} System (\ref{e325})
has at most two solutions in real numbers.
\end{claim}
\proof Omitted.
\\

 In the general case where $\alpha\not=0$ and $\beta\not=0$ from (\ref{e26}) we have $y=\alpha x/(Cx-\beta)$. Hence,  system (\ref{e325})
reduces to the following polynomial equation of degree $n+m$
\be{e126}
x^n(\alpha x)^m=\kappa_1(Cx- \beta)^m.
\ee
We solve (\ref{e126}) using Mathematica and find all its real solutions. We note that if  $n+m$ is odd then the solution is unique, otherwise (\ref{e126}) may have two solutions.

So far we have found all admissible values for the unknown $x=b_1-a_1$ and $y=b_2-a_2$. We solve in an analogous way the remaining three vector equations of (\ref{k6}) with respect to the unknowns $d_1-a_1$ and $d_2-a_2$; $c_1-d_1$ and $c_2-d_2$; $c_1-b_1$ and $c_2-b_2$, respectively. Recall that the four systems
have same matrix $M$ and differ by their right sides only.

Further, we select all combinations of quadruplets $(x_1,x_2,x_3,x_4)$ and $(y_1,y_2,y_3,y_4)$ such that
$(x_i,y_i)$, $i=1,\dots ,4$ are solutions to the first four equations of (\ref{k6}), respectively, and in addition satisfy the following conditions
\be{e30}
x_1-x_2-x_3+x_4=0,\quad y_1-y_2-y_3+y_4=0.
 \ee
Next we show how to obtain the eight unknowns $a_i,b_i,c_i,d_i$, $i=1,2$, from the selected quadruplets $(x_1,x_2,x_3,x_4)$ and $(y_1,y_2,y_3,y_4)$, if any. Let $(x_1,x_2,x_3,x_4)$ and $(y_1,y_2,y_3,y_4)$ be a couple of the selected quadruplets. Since $b_i,c_i,d_i$, $i=1,2$, can be represented through $a_1,a_2$ as
 \be{e33}
 b_1=a_1+x_1, b_2=a_2+y_1;\ c_1=a_1+x_2+x_3, c_2=a_2+y_2+y_3; d_1=a_1+x_1, d_2=a_2+y_2,
 \ee
then we have to find $a_1,a_2$ only. We replace $(x_i,y_i)$, $i=1,\dots ,4$, and the relations (\ref{e33}) in the four vector equations of (\ref{k6}) and for each of them we obtain a linear equation of $a_1$ and $a_2$. If the system of these four linear equations is consistent, i.~e. its rank is 2,
and in addition the corresponding $a_i,b_i,c_i,d_i$, $i=1,2$, satisfy the last four equations of (\ref{k6}) then $(a_1,a_2,b_1,b_2,c_1,c_2,d_1,d_2)$ is a solution to (\ref{k6}).
Otherwise, the selected couple of quadruplets  does not produce a solution to (\ref{k6}).
In this case, if system (\ref{k6}) is consistent it may have at most two solutions and we have shown how to find both of them. \hfill $\Box$

The following theorem holds.
\begin{thm}\label{thm3}
The irreducible TPB surfaces $S_1$ and $S_2$ of degrees $(n,m)$ and $(n+m,n+m)$, respectively, have coincidence if and only if  system (\ref{k6}) is consistent and the control polygon of the surface corresponding to any solution to (\ref{k6}) coincide with the control polygon of $S_2$ (up to eight different enumeration of the control points).
\end{thm}

\proof $\Rightarrow$ Let $S_1$ and $S_2$ have coincidence. Then according to statement (ii) of Proposition\,\ref{proposition1} there exist eight numbers $a_i,b_i,c_i,d_i\in\mathbb{R}$, $i=1,2$, defining transformation $\psi$ which are a solution to system (\ref{k6}) and $S_2(s,t)=S_1(\psi)$. The control polygon of the surface corresponding to the solution $(a_1,a_2,b_1,b_2,c_1,c_2,d_1,d_2)$ coincides with the control polygon of $S_2$.

$\Leftarrow$ Let system (\ref{k6}) be consistent. Then, according to Lemma~\ref{lem4}, it may have at most two solutions. Since there is a solution such that the control polygon of the surface $S$ corresponding to this solution coincides with the control polygon of $S_2$ (up to eight different enumerations of the control polygons) then $S$ and $S_2$ coincide according to Theorem~1 in \citep{Vlachkova}.
\hfill $\Box$

We continue by providing an efficient approach for testing $S_1$ and $S_2$ for coincidence. First, we find all real solutions to system (\ref{k6}) using the proposed technique in Lemma~\ref{lem4}.
Clearly, if system (\ref{k6}) has no real solution then by Proposition~\ref{proposition1} no transformation $\psi$ exists and $S_1$ and $S_2$ are different. Otherwise, let $(a_1,a_2,b_1,b_2,c_1,c_2,d_1,d_2)$ be a solution to system (\ref{k6}).
Following Theorem~\ref{thm3}, we need to compute the control polygon of surface $S$ corresponding to this solution
and to check if it coincides with the control polygon of $S_2$. If these polygons coincide (up to eight different enumerations of the control points) then $S_1$ and $S_2$ have coincidence, otherwise they are different and we continue by checking the second solution to system (\ref{k6}), if any.

Next we describe how we compute the control points of the corresponding surface $S$ defined in quadrangle $\Box ABCD$ with vertices $A(a_1,b_1)$, $B(b_1,b_2)$, $C(c_1,c_2)$, and $D(d_1,d_2)$ by (\ref{e14}). First, we compute the shared domain $G$ of $S$ and $S_2$.
If $G$ is the empty set then $S_1$ and $S_2$ are disjoint. Otherwise $S_1$ and $S_2$ have coincident part and we find it by using the blossoming principle. In this case, unlike the case where $S_1$ and $S_2$ are of same degree, the shared domain can be a polygon with at most eight vertices, see Fig.~\ref{nm-hexagon}. If the number of the polygon vertices is even  we represent the coincident part as a union of TPB surfaces. For example, the surface in Fig.~\ref{nm-hexagon}{\bf c.} is represented as a union of two TPB surfaces. If the number of the polygon vertices is odd we represent the coincident part as a union of TPB surfaces and a triangular B\'{e}zier (TB) surface.   Similar to TPB surface, we compute the control points of the TB surface using the blossoming principle. In \citep[p. 331]{RG} and \citep{YZ} it is pointed out that for any polynomial surface path $S(u,v)=\sum_{i=0}^n\sum_{j=0}^m{\bf c}_{ij}u^iv^j$ of total degree $n+m$ defined in a triangle $\triangle MNP$, the
 B\'{e}zier control points ${\bf q}_{\nu\mu}$ of this surface patch are
\be{e31}
{\bf q}_{\nu\mu}=b^{\triangle} (\underbrace{M,\dots ,M}_{\nu},\underbrace{N,\dots ,N}_{\mu},\underbrace{P,\dots P}_{n+m-\nu-\mu}),
\ee
where $b^{\triangle}((u_1,v_1),\dots ,(u_{n+m},v_{n+m}))=\sum_{i=0}^n\sum_{j=0}^m{\bf c}_{ij}b_{ij}^{\triangle}((u_1,v_1),\dots ,(u_{n+m},v_{n+m}))$
is the blossom of $S(u,v)$, and
\be{e32}
b_{ij}^{\triangle} ((u_1,v_1),\dots ,(u_{n+m},v_{n+m}))=\sum_{\begin{array}{c}{\scriptstyle\{\alpha_1,\dots ,\alpha_i\}\subseteq K}\\ {\scriptstyle\{\beta_1,\dots ,\beta_j\}\subseteq K\backslash\{\alpha_1,\dots ,\alpha_i\}}\end{array}}
\frac{u_{\alpha_1}\dots u_{\alpha_i}v_{\beta_1}\dots v_{\beta_j}}{\binom{n+m}{i,j}},
\ee
$K=\{1,\dots ,n+m\}$,  $\binom{n+m}{i,j}=\frac{(n+m)!}{i!j!(n+m-i-j)!}$,
 is the blossom  of the monomial $u^iv^j$.
  In the case where either $i=0$, or $j=0$ the corresponding sum in (\ref{e32}) equals 1. In the next corollary we present (\ref{e31}) in an equivalent closed form that is more suitable for evaluations of the control points.

 \begin{cor}\label{cor2}
 B\'{e}zier control points ${\bf q}_{\nu\mu}$, $\nu=0,\dots ,n$, $\mu=0,\dots ,m$,defined by (\ref{e31}) are
 \be{e300}
 {\bf q}_{\nu\mu}=\sum_{i=0}^n\sum_{j=0}^m\frac{{\bf c}_{ij}}{\binom{n+m}{i,j}} b_{ij}^{\triangle}= \sum_{j_{\beta}=\underline{j}_{\beta}}^{\bar{j}_{\beta}}
 \sum_{j_{\alpha}=\underline{j}_{\alpha}}^{\bar{j}_{\alpha}}
 \sum_{i_{\beta}=\underline{i}_{\beta}}^{\bar{i}_{\beta}}
 \sum_{i_{\alpha}=\underline{i}_{\alpha}}^{\bar{i}_{\alpha}}
 \binom{\nu}{i_{\alpha}}
 \binom{\mu}{i_{\beta}}
 \binom{\lambda}{i_{\gamma}}
 \binom{\nu -i_{\alpha}}{j_{\alpha}}
 \binom{\mu -i_{\beta}}{j_{\beta}}
 \binom{\lambda -i_{\gamma}}{j_{\gamma}}
 a_1^{i_{\alpha}}b_1^{i_{\beta}}c_1^{i_{\gamma}}
 a_2^{j_{\alpha}}b_2^{j_{\beta}}c_2^{j_{\gamma}},
 \ee
where
 $$
 \begin{array}{l}
\lambda=n+m-\nu -\mu ,\  i_{\gamma}=i-i_{\alpha}-i_{\beta},\ j_{\gamma}=j-j_{\alpha}-j_{\beta},\\
 \underline{i}_{\alpha}=\max (0,i-\mu-\lambda),\ \bar{i}_{\alpha}=\min (i,\nu),\\
 \underline{i}_{\beta}=\max (0,i-\nu-\lambda),\ \bar{i}_{\beta}=\min (i-i_{\alpha},\mu),\\
 \underline{j}_{\alpha}=\max (0,j-(\mu-i_{\beta})-(\lambda - i_{\gamma})),\  \bar{j}_{\alpha}=\min (j,\nu -i_{\alpha}),\\
  \underline{j}_{\beta}=\max (0,j-(\nu-i_{\alpha})-(\lambda - i_{\gamma})),\  \bar{j}_{\beta}=\min (j-j_{\alpha},\mu -i_{\beta}).
\end{array}
 $$
\end{cor}

We outline our procedure for testing $S_1$ and $S_2$ for coincidence in algorithmic form in Algorithm~\ref{alg2}.
\begin{algorithm}[ht]
\caption{Testing two irreducible TPB surfaces of degree $(n,m)$ and $(n+m,n+m)$ for coincidence}\label{alg2}
{\small
\begin{tabular}{rl}
  {\sl Input:}& Irreducible TPB surfaces $S_1$ and $S_2$ of degree $(n,m)$ and $(n+m,n+m)$ given by their control polygons\\
    {\sl Output:}& \ \ (i) $S_1$ and $S_2$ are different;\\
                & \ (ii) $S_1$ and $S_2$ are disjoint;\\
                & (iii) $S_1$ and $S_2$ have coincident part. Report its control\\
                &points.\\
   {\sl Step~1.}&Compute vectors $\boldsymbol{\rho}$, ${\boldsymbol{\rho}}^{1,0}$, ${\boldsymbol{\rho}}^{0,1}$, ${\boldsymbol{\delta}}_i$, and ${\boldsymbol{\delta_i}}^1$, $i=1,\dots ,4$.\\
   {\sl Step~2.}& {\bf If} $\boldsymbol{\rho}$ and ${\boldsymbol{\delta}}_i$ for any $i$, $i=1,\dots ,4$, are non-collinear  \\
                &\quad {\bf then} return (i);\\
                &\qquad {\bf else} compute $\kappa_i$ such that $\kappa_i\boldsymbol{\rho}=\boldsymbol{\delta}_i$, $i=1,\dots ,4$.\\
   {\sl Step~3.}& Solve the first four vector equations of (\ref{k6}) according to Case~1 and Case~2 of Lemma~\ref{lem4}\\
   & and find all admissible couples of quadruplets $(x_1,x_2,x_3,x_4)$ and $(y_1,y_2,y_3,y_4)$.\\
   {\sl Step~4.}& For any admissible couple of quadruplets\\
                & {\bf if} either $x_1-x_2-x_3+x_4\not=0$, or $y_1-y_2-y_3+y_4\not=0$           \\
                &\quad {\bf then} return (i);\\
                &\qquad {\bf else} {\bf if} $x_i^ny_i^m\not=k_i$ for any $i$, $i=1,\dots ,4$   \\
                & \hspace{.48cm}\qquad\quad {\bf then} return (i);\\
                & \hspace{.48cm} \qquad\qquad {\bf else} substitute (\ref{e33}) in the vector equations of (\ref{k6}) and\\
                &\hspace{.96cm}\qquad\qquad  obtain a system of four linear equations of $a_1$ and $a_2$.\\
                &\hspace{.96cm}\qquad\qquad  {\bf If} this system is inconsistent\\
                &\hspace{.96cm}\qquad\qquad \quad{\bf then} return (i);\\
                &\hspace{.96cm}\qquad\qquad \qquad {\bf else} $(a_1,a_2,b_1,b_2,c_1,c_2,d_1,d_2)$ is a solution to (\ref{k6}).\\
     {\sl Step~5.}& Compute the control polygon of the transformed TPB surface $S(s,t)=S_1(\psi(s,t))$ using (\ref{e226}) and (\ref{e300})\\
   &and compare it to the control polygon of $S_2$.\\
    &{\bf If} they coincide (up to eight different enumerations)\\
    &\quad {\bf then} $S_1$ and $S_2$ have coincidence;\\
    &\qquad {\bf else} return (i).\\
    {\sl Step~6.}& Compute the shared domain $G$ of $S$ and $S_2$.\\
   &{\bf If} $D=\emptyset$\\
   &\quad {\bf then} return (ii);\\
   &\qquad {\bf else} divide $G$ into quadrangles and a triangle (if necessary), compute the control points of\\
   &\hspace{.58cm}\qquad the corresponding  coincident \bez surfaces using (\ref{e226}) and (\ref{e300}), and return (iii).
 \end{tabular}
 }
  \end{algorithm}

\begin{rmk}\label{rmk3}
In the case where the shared domain is not a rectangle, multiple representations of the  coincident part as a union of \bez surfaces exist.
\end{rmk}

\begin{rmk}\label{rmk2}
The case where $S_1$ is of degree $(n,m)$ and $S_2$ is of degree either $(n,n+m)$, or $(n+m,m)$ is analogous to the case where $S_2$ is of degree $(n+m,n+m)$. The only difference is that (\ref{e16}) becomes (e.\,g. for degree $(n+m,m)$)
$$\sum_{i=0}^{n}\sum_{j=0}^{m}{\bf p}^1_{ij}B_i^n(u)B_j^m(v)=\sum_{i=0}^{n+m}\sum_{j=0}^{m}{\bf p}^2_{ij}B_i^{n+m}(s)B_j^{m}(t).
$$
Further, the arguments are the same as in the case where $S_1$ is of degree $(n,m)$ and $S_2$ is of degree $(n+m,n+m)$.
\end{rmk}

\section{Examples and results}
\label{sec:2}

We have implemented our method using Mathematica package. In this section we present and analyze the results from our experimental work. In our examples we consider irreducible TPB surfaces.

\begin{example}\label{example1}
This example illustrates the only case where the direct generalization of the algorithm for curves works, see \citep{Vlachkova}. More precisely, this is the case where the two TPB surfaces of same degree have overlapping boundary curves. Here we apply the new method outlined in Algorithm~\ref{alg1}. The irreducible surfaces $S_1$ and $S_2$ are of degree $(4,2)$. Their control points are shown in Table~\ref{table1}. The unique solution of system (\ref{k2}) is $(a^*,b^*,c^*,d^*)=(0,1/2,1/6,3/4)$. Surfaces $S_1$ and $S_2$ are shown in Fig.~\ref{nm1}{\bf a.} and Fig.~\ref{nm1}{\bf b.}, respectively. Their coincident part is surface $S_2$. Both surfaces and their coincident part with its control polygon are shown  in Fig.~\ref{nm2}{\bf a.}
\begin{center}
\begin{figure}[htbp]
	\begin{minipage}[b]{1.5in}
		\hspace*{1cm}
		\includegraphics[width=1.2\textwidth]{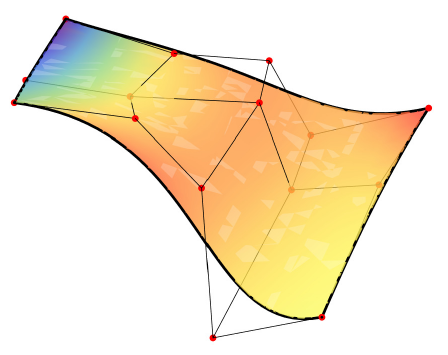}
		\hspace*{1.5cm}{\small {\bf a.}}
	\end{minipage}
\hspace{.5cm}
	~~~~~~~~~
	\begin{minipage}[b]{1.5in}
		\hspace*{1cm}
		\includegraphics[width=1.2\textwidth]{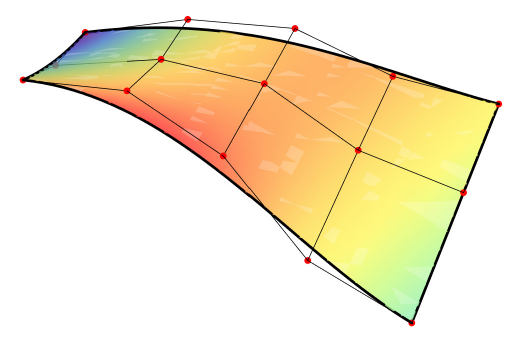}
		\hspace*{1.5cm}{\small {\bf b.}}
	\end{minipage}
\hspace{.5cm}
	~~~~~~~~~
\begin{minipage}[b]{1.5in}
	\hspace*{1cm}
	\includegraphics[width=1.\textwidth]{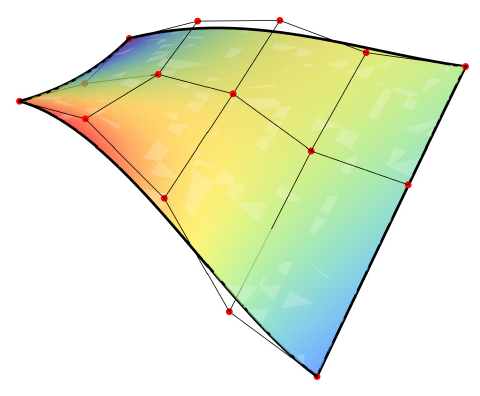}
	\hspace*{1.5cm}{\small {\bf c.}}
\end{minipage}
	\caption{\small  (illustrates examples~\ref{example1} and \ref{example2}) Three irreducible TPB surfaces of degree $(4,2)$ with control points shown in Table~\ref{table1}. {\bf a.} Surface $S_1$; {\bf b.} Surface $S_2$; {\bf c.} Surface $S_3$;
			}\label{nm1}
\end{figure}
\begin{figure}[hbtp]
	\begin{minipage}[b]{1.85in}
		\hspace*{1cm}
		\includegraphics[width=1.55\textwidth]{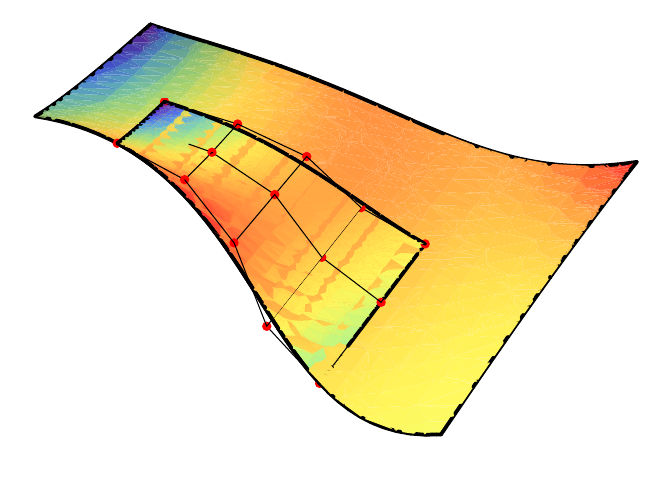}
		\hspace*{1.5cm}{\small {\bf a.}}
	\end{minipage}
	\hspace{1.5cm}
	~~~~~~~~~
	\begin{minipage}[b]{1.85in}
		\hspace*{1cm}
		\includegraphics[width=1.55\textwidth]{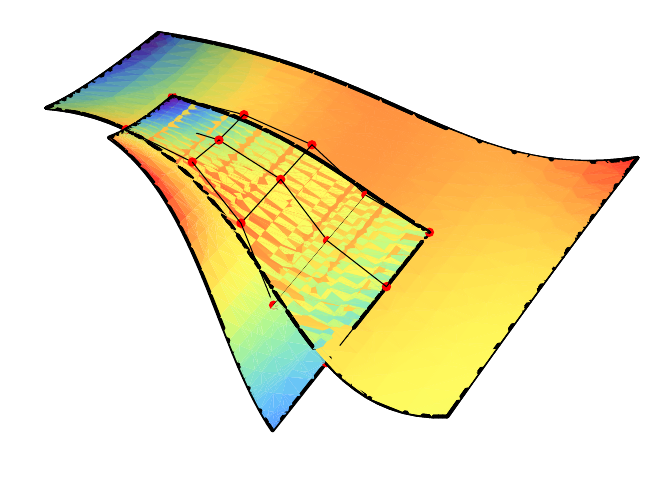}
		\hspace*{1.5cm}{\small {\bf b.}}
	\end{minipage}
	\caption{\small Testing for coincidence the surfaces shown in Fig.~\ref{nm1}. {\bf a.} Example \ref{example1}: The coincident part of $S_1$ and $S_2$ is $S_2$;
			{\bf b.} Example \ref{example2}: The coincident part of $S_1$ and $S_3$ is $S_2$.
		}\label{nm2}
\end{figure}
\end{center}

\begin{table}[htbp]
\caption{\label{table1} The control points of irreducible TPB surfaces $S_1$, $S_2$, $S_3$ of degree $(4,2)$ from examples \ref{example1} and \ref{example2}}
\centering
\begin{tabular}{lllll}
\hline
\noalign{\smallskip}
$S_1$ &$ {\bf p}_{00}=(0, 0, 0)$ &$ {\bf p}_{01}=(0, \frac{7}{10}, -\frac{1}{2})$  &$ {\bf p}_{02}=(0, 2, -\frac{4}{5})$  \\[1ex]
& ${\bf p}_{10}=(1, 0, 1)$& ${\bf p}_{11}=(1, 1, 0)$ & ${\bf p}_{12}=(1, 1, 1)$  \\[1ex]
& ${\bf p}_{20}=(2, 0, \frac{3}{4})$ &$ {\bf p}_{21}=(2, 1, \frac{6}{5})$ &$ {\bf p}_{22}=(2, 2, \frac{3}{4})$ \\[1ex]
&$ {\bf p}_{30}=(3, -1, \frac{3}{20})$ & ${\bf p}_{31}=(3, 1, \frac{2}{5})$ & ${\bf p}_{32}=(3, 2, \frac{1}{5})$\\[1ex]
& ${\bf p}_{40}=(4, 0, \frac{1}{5})$ & ${\bf p}_{41}=(4, 2, \frac{1}{4})$ & ${\bf p}_{42}=(4, 3, \frac{3}{4})$\\

\noalign{\smallskip}\hline\noalign{\smallskip}
$S_2$ & ${\bf p}_{00}=(\frac{2}{3}, -\frac{5}{324}, \frac{6157}{12960})$ & ${\bf p}_{01}=(\frac{2}{3}, \frac{2179}{5184}, \frac{9829}{51840})$ & ${\bf p}_{02}=(\frac{2}{3}, \frac{1073}{1296}, \frac{9689}{103680})$  \\[1ex]
& ${\bf p}_{10}=(\frac{5}{4}, \frac{-23}{432}, \frac{425}{576})$ & ${\bf p}_{11}=(\frac{5}{4}, \frac{1567}{3456}, \frac{5881}{11520})$ & ${\bf p}_{12}=(\frac{5}{4}, \frac{1447}{1728}, \frac{6673}{13824})$   \\[1ex]
&$ {\bf p}_{20}=(\frac{11}{6}, -\frac{1}{6}, \frac{3907}{5760})$ &$ {\bf p}_{21}=(\frac{11}{6}, \frac{107}{256}, \frac{15241}{23040})$ &$ {\bf p}_{22}=(\frac{11}{6}, \frac{515}{576}, \frac{30277}{46080})$  \\[1ex]
&$ {\bf p}_{30}=(\frac{29}{12}, -\frac{27}{64}, \frac{1837}{3840})$ &$ {\bf p}_{31}=(\frac{29}{12}, \frac{165}{512}, \frac{2837}{5120})$  &$ {\bf p}_{32}=(\frac{29}{12}, \frac{703}{768}, \frac{17401}{30720})$  \\[1ex]
&$ {\bf p}_{40}=(3, -\frac{27}{64}, \frac{849}{2560})$ &$ {\bf p}_{41}=(3, \frac{2287}{5120}, \frac{4253}{10240})$  &$ {\bf p}_{42}=(3, \frac{1433}{1280}, \frac{9489}{20480})$\\

\noalign{\smallskip}\hline\noalign{\smallskip}
$S_3$ &$ {\bf p}_{00}=(\frac{2}{3}, -\frac{11941}{32400}, \frac{475469}{648000})$ &$ {\bf p}_{01}=(\frac{2}{3}, \frac{247}{960}, \frac{29557}{129600})$  &$ {\bf p}_{02}=(\frac{2}{3}, \frac{1073}{1296}, \frac{9689}{103680})$ \\[1ex]
&$ {\bf p}_{10}=(\frac{5}{4}, -\frac{3443}{7200}, \frac{411113}{432000})$ &${\bf p}_{11}= (\frac{5}{4}, \frac{1727}{5760}, \frac{11267}{21600})$ &$ {\bf p}_{12}=(\frac{5}{4}, \frac{1447}{1728}, \frac{6673}{13824})$ \\[1ex]
&$ {\bf p}_{20}=(\frac{11}{6}, -\frac{1877}{2880}, \frac{199789}{288000})$ &$ {\bf p}_{21}=(\frac{11}{6}, \frac{2621}{11520}, \frac{849}{1280})$ &$ {\bf p}_{22}=(\frac{11}{6}, \frac{515}{576}, \frac{30277}{46080})$ \\[1ex]
&$ {\bf p}_{30}=(\frac{29}{12}, -\frac{9997}{9600}, \frac{26091}{64000})$ &$ {\bf p}_{31}=(\frac{29}{12}, \frac{653}{7680}, \frac{659}{1200})$ &$ {\bf p}_{32}=(\frac{29}{12}, \frac{703}{768}, \frac{17401}{30720})$\\[1ex]
&$ {\bf p}_{40}=(3, -\frac{36737}{32000}, \frac{33149}{128000})$ &$ {\bf p}_{41}=(3, \frac{909}{5120}, \frac{10141}{25600})$ &$ {\bf p}_{42}=(3, \frac{1433}{1280}, \frac{9489}{20480})$\\
\noalign{\smallskip}\hline\noalign{\smallskip}
\end{tabular}
\end{table}
\end{example}

\begin{example}\label{example2}
We test for coincidence the irreducible surfaces $S_1$ and $S_3$ of degree $(4,2)$ with control points shown in Table~\ref{table1}. We apply Algorithm~\ref{alg1}. The unique solution of system (\ref{k2}) is $(a^*,b^*,c^*,d^*)=(-1/5,1/2,1/6,3/4)$. Surfaces $S_1$ and $S_3$ are shown in Fig.~\ref{nm1}{\bf b.} and Fig.~\ref{nm1}{\bf c.}, respectively. Their coincident part is $S_2$ and it is shown with its control polygon in Fig.~\ref{nm1}{\bf b.}
\end{example}

\begin{example}\label{example3}
We consider the irreducible surfaces $S_1$ and $S_2$ of degree $(2,3)$ and $(5,5)$, respectively. Their control points are shown in Table~\ref{table2}. This example matches Case 1 of Lemma~\ref{lem4},  and hence system (\ref{k6}) has a unique solution
$(a_1,a_2,b_1,b_2,c_1,c_2,d_1,d_2)$=$(0,1/2,1/2,0,1,1/2,1/2,1)$. Surfaces $S_1$ and $S_3$ are shown in Fig.~\ref{nm-3-2}{\bf a.} and Fig.~\ref{nm-3-2}{\bf b.}, respectively. They have coincidence and their coincident part is surface $S_2$, see Fig.~\ref{nm-3-2}{\bf c.}

\begin{figure}[htbp]
	\centering
	\begin{minipage}[b]{1.3in}
			\includegraphics[width=1.\textwidth]{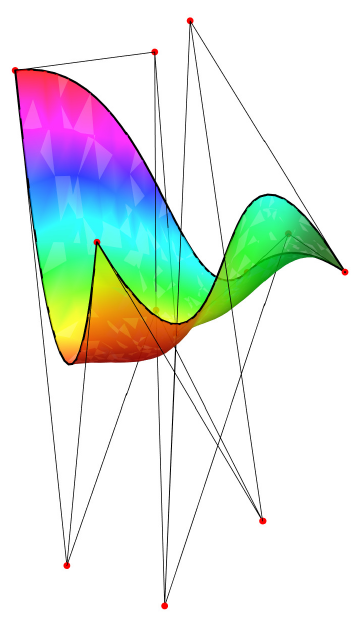}
			{\small {\bf a.}}
	\end{minipage}
	~~~~~~~~~
		\begin{minipage}[b]{1.5in}
				\includegraphics[width=1.2\textwidth]{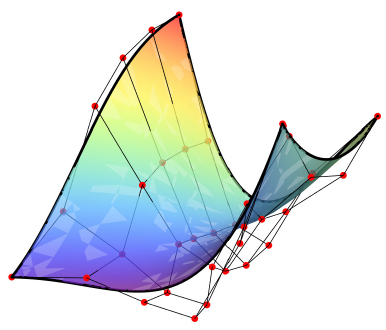}
				\vspace*{1cm}
{\small {\bf b.}}
	\end{minipage}
~~~~~~~~~
\begin{minipage}[b]{1.2in}
		\includegraphics[width=1.2\textwidth]{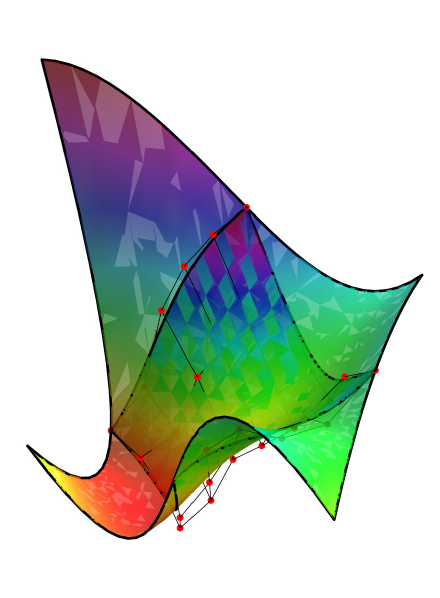}
		{\small {\bf c.}}
	\vspace*{1cm}
\end{minipage}
\caption{\small The irreducible TPB surfaces $S_1$ of degree $(2,3)$ and $S_2$  of degree $(5,5)$ from Example~\ref{example3}. The corresponding control points are shown in Table~\ref{table2}. {\bf a.} Surface $S_1$;  {\bf b.} Surface $S_2$; {\bf c.} The coincident part of $S_1$ and $S_2$ is $S_2$.}\label{nm-3-2}
\end{figure}

\begin{table*}[htbp]
	\caption{\label{table2} The control points of irreducible TPB surfaces of degree $(2,3)$ and $(5,5)$ from Example~\ref{example3}}
			\centering
		\begin{tabular}{lllll}
			\hline
			\noalign{\smallskip}
			$S_1$ &$ {\bf p}_{00}=(0, 0, 0)$
			& ${\bf p}_{01}=(\frac{3}{2}, 0, -3)$
			&$ {\bf p}_{02}=(\frac{3}{2}, 0, 3)$
			&${\bf p}_{03}=(3, 0, 0)$\\[1ex]
			& ${\bf p}_{10}=(0, \frac{3}{2}, -3)$
			&${\bf p}_{11}= (\frac{4}{3}, \frac{4}{3}, 0)$
			&$ {\bf p}_{12}=(\frac{7}{6}, \frac{5}{3}, -\frac{10}{3})$
			&${\bf p}_{13}=(3, 1, 1)$\\[1ex]
			&${\bf p}_{20}=(0, \frac{3}{2}, 3)$  &$ {\bf p}_{21}=(\frac{5}{3}, \frac{7}{6}, \frac{10}{3})$&$ {\bf p}_{22}=(\frac{11}{6}, \frac{11}{6}, 0)$
			&${\bf p}_{23}=(3, 2, 1)$\\
			\noalign{\smallskip}\hline\noalign{\smallskip}
			$S_2$ & $ {\bf p}_{00}=(0, \frac{9}{8}, -\frac{3}{4})$
			&$ {\bf p}_{01}=(\frac{7}{16}, \frac{49}{40}, -\frac{1}{5})$
			&$ {\bf p}_{02}=(\frac{31}{40}, \frac{13}{10}, \frac{11}{16})$
			&$ {\bf p}_{03}=(\frac{137}{128}, \frac{879}{640}, \frac{187}{160})$ \\[1ex]
			&$ {\bf p}_{04}=(\frac{11}{8}, \frac{29}{20}, \frac{3}{2})$
			&$ {\bf p}_{05}=(\frac{27}{16}, \frac{25}{16}, \frac{7}{4})$
			
			&$ {\bf p}_{10}=(\frac{7}{16}, \frac{37}{40}, -\frac{4}{5})$
			&$ {\bf p}_{11}=(\frac{157}{200}, \frac{217}{200}, -\frac{27}{50})$ \\[1ex]
			&$ {\bf p}_{12}=(\frac{3367}{3200}, \frac{3901}{3200}, \frac{49}{800})$
			& $ {\bf p}_{13}=(\frac{2073}{1600}, \frac{431}{320}, \frac{13}{40})$
			&$ {\bf p}_{14}=(\frac{1251}{800}, \frac{1177}{800}, \frac{21}{40})$
			&$ {\bf p}_{15}=(\frac{37}{20}, \frac{13}{8}, \frac{7}{10})$\\[1ex]
			&$ {\bf p}_{20}=(\frac{3}{4}, \frac{3}{4}, -\frac{81}{80})$
			&$ {\bf p}_{21}=(\frac{3287}{3200}, \frac{3053}{3200}, -\frac{679}{800})$
			&$ {\bf p}_{22}=(\frac{199}{160}, \frac{181}{160}, -\frac{163}{400})$
			&$ {\bf p}_{23}=(\frac{4659}{3200}, \frac{4149}{3200}, -\frac{223}{800})$ \\[1ex]
			&$ {\bf p}_{24}=(\frac{171}{100}, \frac{29}{20}, -\frac{59}{400})$
			&$ {\bf p}_{25}=(\frac{1277}{640}, \frac{1039}{640}, \frac{3}{160})$
			
			&$ {\bf p}_{30}=(\frac{637}{640}, \frac{351}{640}, -\frac{189}{160})$
			&$ {\bf p}_{31}=(\frac{1957}{1600}, \frac{251}{320}, -\frac{197}{200})$  \\[1ex]
			& ${\bf p}_{32}=(\frac{903}{640}, \frac{3173}{3200}, -\frac{99}{160})$
			&$ {\bf p}_{33}=(\frac{517}{320}, \frac{1887}{1600}, -\frac{14}{25})$
			&$ {\bf p}_{34}=(\frac{6023}{3200}, \frac{4301}{3200}, -\frac{343}{800})$
			&${\bf p}_{35}=(\frac{2703}{320}, \frac{97}{64}, -\frac{7}{40})$\\[1ex]
			
			&$ {\bf p}_{40}=(\frac{99}{80}, \frac{23}{80}, -\frac{3}{4})$
			&$ {\bf p}_{41}=(\frac{1149}{800}, \frac{439}{800}, -\frac{101}{200})$
			&$ {\bf p}_{42}=(\frac{647}{400}, \frac{311}{400}, -\frac{87}{400})$
			&$ {\bf p}_{43}=(\frac{5887}{3200}, \frac{125}{128}, -\frac{183}{800})$ \\[1ex]
			&$ {\bf p}_{44}=(\frac{859}{400}, \frac{91}{80}, -\frac{7}{100})$
			&$ {\bf p}_{45}=(\frac{101}{40}, \frac{103}{80}, \frac{7}{20})$
			
			&$ {\bf p}_{50}=(\frac{3}{2}, 0, 0)$
			&$ {\bf p}_{51}=(\frac{27}{16}, \frac{23}{80}, \frac{3}{20})$  \\[1ex]
			&$ {\bf p}_{52}=(\frac{241}{128}, \frac{343}{640}, \frac{7}{32})$
			&$ {\bf p}_{53}=(\frac{687}{320}, \frac{237}{320}, \frac{1}{40})$
			&$ {\bf p}_{54}=(\frac{101}{40}, \frac{71}{80}, \frac{3}{20})$
			&$ {\bf p}_{55}=(3, 1, \frac{3}{4})$ \\
			\noalign{\smallskip}\hline\noalign{\smallskip}
			
		\end{tabular}
		
\end{table*}

\end{example}

\begin{example}\label{example6}
We test for coincidence the irreducible surfaces $S_1$ and $S_2$ of degree $(2,2)$ and $(4,4)$, respectively. Their control points are shown in Table~\ref{table5}. The corresponding vectors $\boldsymbol{\rho},\boldsymbol{\rho}^{1,0},\boldsymbol{\rho}^{0,1}$ are coplanar and hence, matrix $M$, defined by (\ref{m1}) has rank 2. We apply Algorithm~\ref{alg2} and obtain the following two solutions $(a_1,a_2,b_1,b_2,c_1,c_2,d_1,d_2)$ to system (\ref{e22}), 
$$
\begin{array}{ll}
&(-1,1/3,4/3,1/2,3/4,3/2,-1/4,13/12),\nonumber \\
&{\small(34/71,271/183,75/284, -181/549, -291/284,15/122,-277/568,329/366).\nonumber
}
\end{array}
$$
The first of these solutions generates TPB surface whose control polygon coincides with the control polygon of $S_2$. Hence, $S_1$ and $S_2$ have coincidence. Surfaces $S_1$, $S_2$, and their coincident part $S$ are shown in Fig.~\ref{example4} The control points of $S$ are shown in Table~\ref{table5}.
\begin{center}
\begin{figure}[htbp]
	\centering
	\begin{minipage}[b]{1.5in}
			\includegraphics[width=1.\textwidth]{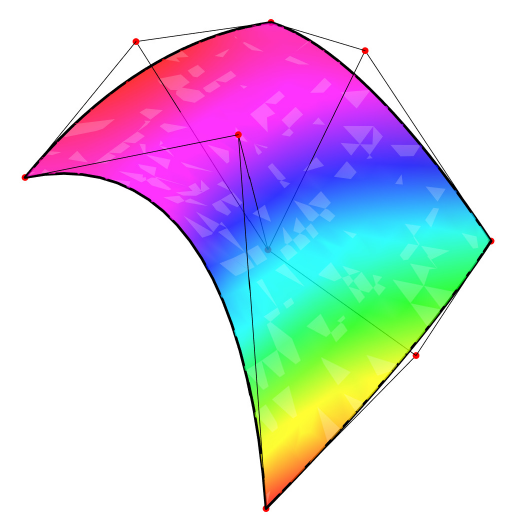}
		\hspace*{1.5cm}{\small {\bf a.}}
	\end{minipage}
	~~~~~~~~~
	\begin{minipage}[b]{1.7in}
		\hspace*{-.4cm}
		\includegraphics[width=1.3\textwidth]{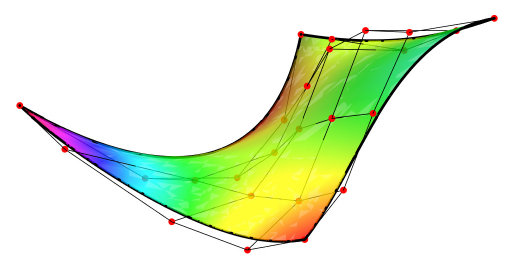}
			\hspace*{1.5cm}{\small {\bf b.}}
	\end{minipage}
		~~~~~~~~~
	\begin{minipage}[b]{1.7in}
		\hspace*{-.4cm}
		\includegraphics[width=1.3\textwidth]{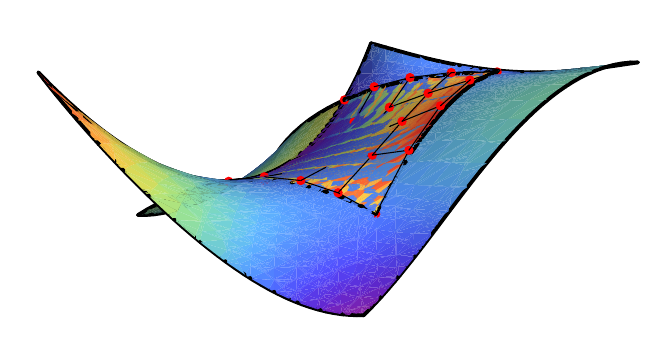}
		\hspace*{1.5cm}{\small {\bf c.}}
		\end{minipage}
	\caption{\small The irreducible TPB surfaces $S_1$ of degree $(2,2)$ and $S_2$  of degree $(4,4)$ and their coincident part $S$ from Example~\ref{example6}. The corresponding control points are shown in Table~\ref{table5}. {\bf a.} Surface $S_1$; {\bf b.} Surface $S_2$; {\bf c.} Surfaces $S_1$, $S_2$, and $S$ which is shown with its control polygon.}\label{example4}
\end{figure}
\end{center}

\begin{table*}
\caption{\label{table5} The control points of irreducible TPB surfaces of degree $(2,2)$ and $(4,4)$ from Example~\ref{example6}}
{
\centering
\begin{tabular}{lllll}
\hline\noalign{\smallskip}
$S_1$ & ${\bf p}_{00}=(0, 0, 0)$
&${\bf p}_{01}=(1, 0, 1)$
& ${\bf p}_{02}=(2, 0, \frac{3}{4}) $&\\[1ex]
&${\bf p}_{10}=(0, \frac{7}{10}, -\frac{1}{2}) $
&$ {\bf p}_{11}=(1, 1, 0)$
&$ {\bf p}_{12}=(2, 1, \frac{6}{5})$ &\\ [1ex]
&${\bf p}_{20}=(0, 2, -\frac{4}{5})$
&${\bf p}_{21}=(1, 1, 1) $
& $ {\bf p}_{22}=(2, 2, \frac{3}{4})$& \\
\noalign{\smallskip}\hline\noalign{\smallskip}

$S_2$ &$ {\bf p}_{00}=(\frac{2}{3}, -\frac{86}{45},\frac{95}{36})$
&$ {\bf p}_{01}=(\frac{25}{24}, -\frac{35}{18}, \frac{409}{180})$
&$ {\bf p}_{02}=(\frac{17}{12}, -\frac{121}{90}, \frac{5951}{5760})$
&$ {\bf p}_{03}=(\frac{43}{24}, -\frac{10109}{11520}, \frac{8609}{23040})$ \\[1ex]
&$ {\bf p}_{04}=(\frac{13}{6}, -\frac{1123}{2304}, \frac{5957}{23040})$

&$ {\bf p}_{10}=(\frac{3}{4}, -\frac{85}{108}, \frac{2027}{2160})$
 &$ {\bf p}_{11}=(\frac{37}{32}, -\frac{93}{160}, \frac{9631}{11520})$
 & ${\bf p}_{12}=(\frac{25}{16}, -\frac{5743}{17280}, \frac{9151}{17280})$ \\[1ex]
 &$ {\bf p}_{13}=(\frac{63}{32}, -\frac{23423}{138240}), \frac{116701}{276480})$
 &$ {\bf p}_{14}=(\frac{19}{8}, \frac{53}{7680}, \frac{23203}{46080})$

&$ {\bf p}_{20}=(\frac{5}{6}, \frac{1051}{2430}, -\frac{1273}{38880})$
&$ {\bf p}_{21}=(\frac{61}{48}, \frac{87029}{155520}, \frac{89543}{311040})$ \\[1ex]
&$ {\bf p}_{22}=(\frac{41}{24}, \frac{427381}{933120}, \frac{1410439}{1866240})$
&$ {\bf p}_{23}=(\frac{103}{48}, \frac{3137}{7680}, \frac{5157}{5120})$
&$ {\bf p}_{24}=(\frac{31}{12}, \frac{1151}{2560}, \frac{154297}{138240})$

&$ {\bf p}_{30}=(\frac{11}{12}, \frac{317}{240}, \frac{101}{2160})$\\[1ex]
&$ {\bf p}_{31}=(\frac{133}{96}, \frac{5191}{4320}, \frac{22847}{34560})$
&$ {\bf p}_{32}=(\frac{89}{48}, \frac{108571}{103680}, \frac{18113}{13824})$
&$ {\bf p}_{33}=(\frac{223}{96}, \frac{5179}{5120}, \frac{150163}{92160})$
&$ {\bf p}_{34}=(\frac{67}{24}, \frac{259}{256}, \frac{12917}{7680})$ \\[1ex]

&$ {\bf p}_{40}=(1, \frac{83}{45}, \frac{63}{80})$
&$ {\bf p}_{41}=(\frac{3}{2}, \frac{947}{576}, \frac{5461}{5760})$
&$ {\bf p}_{42}=(2, \frac{12431}{5760}, \frac{18029}{34560})$
&$ {\bf p}_{43}=(\frac{5}{2}, \frac{4631}{1920}, \frac{2587}{3840})$ \\[1ex]
&$ {\bf p}_{44}=(3, \frac{741}{320}, \frac{621}{640})$&&&  \\
\noalign{\smallskip}\hline\noalign{\smallskip}

$S$ &$ {\bf p}_{00}=(\frac{17}{21}, 0, \frac{4267}{7056})$
&$ {\bf p}_{01}=(\frac{31}{28}, 0, \frac{10609}{14112})$
&$ {\bf p}_{02}=(\frac{59}{42}, 0, \frac{34927}{42336})$
&$ {\bf p}_{03}=(\frac{143}{84}, 0, \frac{277}{336})$ \\[1ex]
&$ {\bf p}_{04}=(2, 0, \frac{3}{4})$

&$ {\bf p}_{10}=(\frac{71}{84}, \frac{1051}{2352}, \frac{11633}{35280})$
&$ {\bf p}_{11}=(\frac{127}{112}, \frac{2227}{4704}, \frac{71797}{141120})$
& ${\bf p}_{12}=(\frac{239}{168}, \frac{6931}{14112}, \frac{143093}{211680})$ \\[1ex]
&$ {\bf p}_{13}=(\frac{575}{336}, \frac{1}{2}, \frac{5587}{6720})$
&$ {\bf p}_{14}=(2, \frac{1}{2}, \frac{39}{40})$

&$ {\bf p}_{20}=(\frac{37}{42}, \frac{4537}{5292}, \frac{9115}{42336})$
&$ {\bf p}_{21}=(\frac{65}{56}, \frac{1163}{1323}, \frac{182957}{423360})$ \\[1ex]
&$ {\bf p}_{22}=(\frac{121}{84}, \frac{28897}{31752}, \frac{817367}{1270080})$
&$ {\bf p}_{23}=(\frac{289}{168}, \frac{479}{504}, \frac{8563}{10080})$
&$ {\bf p}_{24}=(2, 1, \frac{21}{20})$

&$ {\bf p}_{30}=(\frac{11}{12}, \frac{109}{90}, \frac{103}{392})$ \\[1ex]
&$ {\bf p}_{31}=(\frac{19}{16}, \frac{6127}{5040}, \frac{4673}{9408})$
&$ {\bf p}_{32}=(\frac{35}{24}, \frac{4783}{3780}, \frac{24463}{35280})$
&$ {\bf p}_{33}=(\frac{83}{48}, \frac{457}{336}, \frac{5731}{6720})$
&$ {\bf p}_{34}=(2, \frac{3}{2}, \frac{39}{40})$  \\[1ex]

&$ {\bf p}_{40}=(\frac{20}{21}, \frac{662}{441}, \frac{991}{2205})$
&$ {\bf p}_{41}=(\frac{17}{14}, \frac{1313}{882}, \frac{5867}{8820})$
&$ {\bf p}_{42}=(\frac{31}{21}, \frac{2074}{1323}, \frac{41659}{52920})$
&$ {\bf p}_{43}=(\frac{73}{42}, \frac{73}{42}, \frac{137}{168})$ \\[1ex]
&$ {\bf p}_{44}=(2, 2, \frac{3}{4})$&&&  \\
\noalign{\smallskip}\hline\noalign{\smallskip}

\end{tabular}
}
\end{table*}

\end{example}

\begin{center}
	\begin{figure}[htbp]
		\centering
		\begin{minipage}[b]{1.5in}
			\hspace*{-1cm}
			\includegraphics[width=1.2\textwidth]{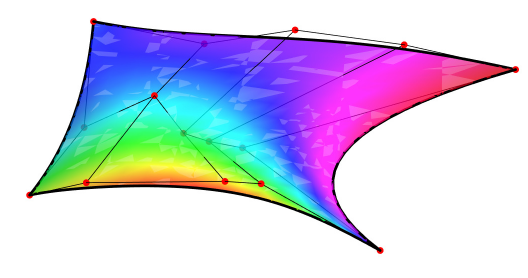}
			\hspace*{1.5cm}{\small {\bf a.}}
		\end{minipage}
		~~~~~~~~~
		\begin{minipage}[b]{1.7in}
			\hspace*{-.9cm}
			\includegraphics[width=1.3\textwidth]{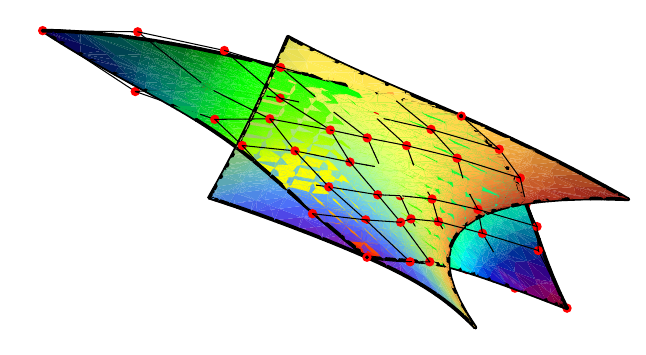}
					\hspace*{1.5cm}{\small {\bf b.}}
		\end{minipage}
				~~~~~~~~~
		\begin{minipage}[b]{1.7in}
			\hspace*{-.4cm}
			\includegraphics[width=1.3\textwidth]{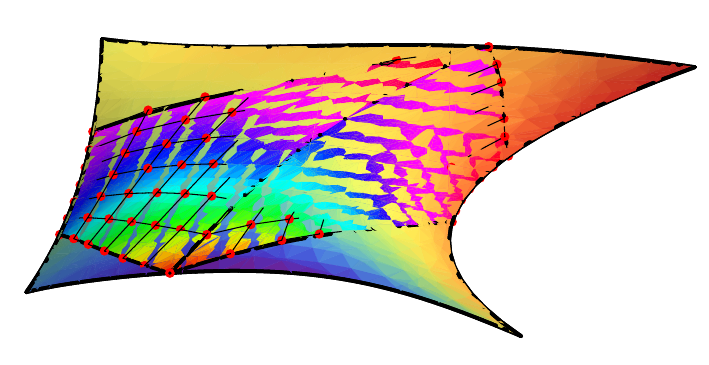}
			\hspace*{1.5cm}{\small {\bf c.}}
				\end{minipage}
		\caption{\small {\bf a.} Irreducible TPB surfaces $S_1$ of degree $(2,4)$ and $S_2$  of degree $(6,6)$ from  Example~\ref{example5}. The control points of $S_1$ and $S_2$ are shown in Table~\ref{table5}. {\bf a.} $S_1$; {\bf b.} $S_1$ and $S_2$ with its control polygon; {\bf c.} The coincident part $S$ is presented as two TPB surfaces.}\label{nm-hexagon}
								\end{figure}
\end{center}

\begin{example}\label{example5}
In our final example we test for coincidence the irreducible surfaces $S_1$ and $S_2$ of degree $(2,4)$ and $(6,6)$, respectively. Their control points are shown in Table~\ref{table3}. Surfaces $S_1$ and $S_2$ are shown in Fig.~\ref{nm-hexagon}{\bf a.} and Fig.~\ref{nm-hexagon}{\bf b.}. The corresponding system (\ref{k6}) has a unique solution $(a_1,a_2,b_1,b_2,c_1,c_2,d_1,d_2)$=$(1/2,-1/2,0,1/3,1/2,7/6,-1,2/3)$. The shared domain of $S_1$ and $S_2$ is a hexagon. We represent the coincident part $S$ as a union of two TPB surfaces and compute their control points using blossoming, see Fig~\ref{nm-hexagon}{\bf c.}
\end{example}

\begin{table*}[hbtp]
\caption{\label{table3} The control points of irreducible TPB surfaces of degree $(2,4)$ and $(6,6)$ from Example~\ref{example5} }
{
\centering
\begin{tabular}{lllll}
\hline\noalign{\smallskip}
$S_1$ & ${\bf p}_{00}=(0, 0, 0)$
&${\bf p}_{01}=(0, \frac{7}{10}, -\frac{1}{2})$
& $ {\bf p}_{02}=(0, 2, -\frac{4}{5})$
&$ {\bf p}_{03}=(0, \frac{23}{10}, -\frac{4}{5})$  \\[1ex]
&$ {\bf p}_{04}=(0, 3, 1)$

&$ {\bf p}_{10}=(1, 0, 1)$
&$ {\bf p}_{11}=(1, 1, 0)$
&$ {\bf p}_{12}=(1, 1, 1)$ \\[1ex]
&$ {\bf p}_{13}=(1, \frac{6}{5}, \frac{6}{5})$
&${\bf p}_{14}=(1, \frac{3}{2}, \frac{4}{3})$

&${\bf p}_{20}=(2, 0, \frac{3}{4})$
&$ {\bf p}_{21}=(2, 1, \frac{6}{5})$\\[1ex]
& $ {\bf p}_{22}=(2, 2, \frac{3}{4})$
&$ {\bf p}_{23}=(2, 3, 1)$
&${\bf p}_{24}=(2, 4, \frac{3}{2})$&\\
\noalign{\smallskip}\hline\noalign{\smallskip}

$S_2$ &${\bf p}_{00}=(1, -\frac{79}{32}, \frac{2717}{768})$
& ${\bf p}_{01}=(\frac{7}{6}, -\frac{4151}{2880}, \frac{19667}{17280})$
& ${\bf p}_{02}=(\frac{4}{3}, -\frac{277}{1800}, \frac{4481}{14400})$
& ${\bf p}_{03}=(\frac{3}{2}, \frac{1361}{3600}, \frac{44467}{64800})$ \\[1ex]

&${\bf p}_{04}=(\frac{5}{3}, \frac{5447}{6075}, \frac{255913}{291600})$
& ${\bf p}_{05}=(\frac{11}{6}, \frac{1991}{1215}, \frac{26399}{29160})$
& ${\bf p}_{06}=(2, \frac{8}{3}, \frac{1687}{1620})$

& ${\bf p}_{10}=(\frac{5}{6}, -\frac{3973}{2880}, \frac{4081}{1728})$ \\[1ex]
&${\bf p}_{11}=(1, -\frac{3143}{4320}, \frac{191717}{207360})$
& ${\bf p}_{12}=(\frac{7}{6}, \frac{2489}{8640}, \frac{250213}{777600})$

& ${\bf p}_{13}=(\frac{4}{3}, \frac{9049}{12150}, \frac{1379747}{2332800})$
& ${\bf p}_{14}=(\frac{3}{2}, \frac{175433}{145800}, \frac{668537}{874800})$ \\[1ex]
&${\bf p}_{15}=(\frac{5}{3}, \frac{13499}{7290}, \frac{307297}{349920})$

& ${\bf p}_{16}=(\frac{11}{6}, \frac{3341}{1215}, \frac{32099}{29160})$

& ${\bf p}_{20}=(\frac{2}{3}, -\frac{2269}{3600}, \frac{22961}{21600})$
& ${\bf p}_{21}=(\frac{5}{6}, -\frac{1883}{14400}, \frac{157141}{388800})$ \\[1ex]

&${\bf p}_{22}=(1, \frac{26249}{38880}, \frac{2976581}{23328000})$
& ${\bf p}_{23}=(\frac{7}{6}, \frac{2027447}{1944000}, \frac{1061101}{2332800})$
& $ {\bf p}_{24}=(\frac{4}{3}, \frac{520417}{364500}, \frac{6107587}{8748000}) $
& ${\bf p}_{25}=(\frac{3}{2}, \frac{96541}{48600}, \frac{264529}{291600})$ \\[1ex]
&${\bf p}_{26}=(\frac{5}{3}, \frac{5609}{2025}, \frac{115331}{97200})$

& ${\bf p}_{30}=(\frac{1}{2}, -\frac{203}{900}, \frac{14257}{32400})$
& ${\bf p}_{31}=(\frac{2}{3}, \frac{55237}{194400}, \frac{96383}{583200})$

& ${\bf p}_{32}=(\frac{5}{6}, \frac{1863737}{1944000}, \frac{17473}{233280})$ \\[1ex]
&${\bf p}_{33}=(1, \frac{272407}{216000}, \frac{4416611}{10368000})$
& ${\bf p}_{34}=(\frac{7}{6}, \frac{3084847}{1944000}, \frac{1652957}{2332800})$
& ${\bf p}_{35}=(\frac{4}{3}, \frac{33577}{16200}, \frac{27907}{28800})$
& ${\bf p}_{36}=(\frac{3}{2}, \frac{89399}{32400}, \frac{246491}{194400})$ \\[1ex]

&${\bf p}_{40}=(\frac{1}{3}, \frac{1819}{12150}, \frac{18443}{145800})$
& ${\bf p}_{41}=(\frac{1}{2}, \frac{12494}{18225}, -\frac{1339}{437400})$

& ${\bf p}_{42}=(\frac{2}{3}, \frac{904483}{729000}, \frac{35749}{4374000})$
&${\bf p}_{43}=(\frac{5}{6}, \frac{318233}{216000}, \frac{29029}{77760})$ \\[1ex]

&${\bf p}_{44}=(1, \frac{1014119}{583200}, \frac{48464383}{69984000})$
& ${\bf p}_{45}=(\frac{7}{6}, \frac{100631}{46656}, \frac{6996317}{6998400})$
& ${\bf p}_{46}=(\frac{4}{3}, \frac{133901}{48600}, \frac{1543601}{1166400})$

& ${\bf p}_{50}=(\frac{1}{6}, \frac{2891}{4860}, -\frac{1411}{7290})$ \\[1ex]
&${\bf p}_{51}=(\frac{1}{3}, \frac{1613}{1458}, -\frac{42221}{174960})$
& ${\bf p}_{52}=(\frac{1}{2}, \frac{37217}{24300}, -\frac{18343}{145800})$

& ${\bf p}_{53}=(\frac{2}{3}, \frac{328277}{194400}, \frac{169903}{583200})$
& ${\bf p}_{54}=(\frac{5}{6}, \frac{2213237}{1166400}, \frac{2383159}{3499200})$  \\[1ex]

&${\bf p}_{55}=(1, \frac{87773}{38880}, \frac{1983853}{1866240})$
& ${\bf p}_{56}=(\frac{7}{6}, \frac{8009}{2880}, \frac{224443}{155520})$

& ${\bf p}_{60}=(0, \frac{17}{15}, -\frac{203}{405})$
& ${\bf p}_{61}=(0, \frac{17}{15}, -\frac{203}{405})$  \\[1ex]
&${\bf p}_{62}=(\frac{1}{3}, \frac{22679}{12150}, -\frac{24437}{145800})$

& ${\bf p}_{63}=(\frac{1}{2}, \frac{31721}{16200}, \frac{34481}{97200})$
& ${\bf p}_{64}=(\frac{2}{3}, \frac{206657}{97200}, \frac{499307}{583200})$
& ${\bf p}_{65}=(\frac{5}{6}, \frac{190249}{77760}, \frac{62737}{46656})$ \\[1ex]

&${\bf p}_{66}=(1, \frac{7553}{2592}, \frac{112733}{62208})$&&& \\[2pt]
\noalign{\smallskip}\hline\noalign{\smallskip}

\end{tabular}
}
\end{table*}

\section{Conclusions and future work}
\label{sec:3}

In this paper we considered and solved the problem for testing TPB surfaces for coincidence. We presented two different methods and develop two algorithms based on these methods that test two irreducible TPB surfaces for coincidence and report their coincident part if it is present. The first algorithm works for surfaces of same degree $(m,n)$ and the second one - for surfaces of degree $(m,n)$ and $(m+n,m+n)$, respectively.
We presented numerical experiments and gave examples to visualize and support the obtained results.
Our next task for future research is to develop and implement an algorithm for testing TB surfaces for coincidence.

\section*{Acknowledgments}

This work was partially supported by Sofia University Science Fund
Grant No. 80-10-103/2023.

\bibliography{CGTA2017bibfile}

\end{document}